\documentclass[pra,floatfix,twocolumn,superscriptaddress,showpacs,preprintnumbers,longbibliography,nofootinbib]{revtex4-1}

\usepackage[utf8]{inputenc} 
\usepackage{bibunits} 
\usepackage{amssymb,amsfonts,amsmath}
\usepackage{float} 
\usepackage{amsthm}
\usepackage{enumitem}
\usepackage{bbold}
\usepackage[normalem]{ulem}
\usepackage{hyperref} 
\hypersetup{colorlinks=true, urlcolor=blue, citecolor=blue,linkcolor=blue}
\usepackage[all]{hypcap} 

\usepackage[mathscr]{euscript}
\usepackage{graphicx}
\usepackage{dcolumn}
\usepackage{bm}
\usepackage{psfrag}
\usepackage{epsfig}
\usepackage{multirow}
\usepackage{color}
\usepackage{bbm}
\usepackage[FIGTOPCAP,raggedright,nooneline]{subfigure}
\usepackage{verbatim}
\usepackage{upgreek} 
\usepackage{physics}
\usepackage{tikz} 

\newcommand{\ignore}[1]{}

\newcommand{\PT}{$\mathcal{PT}$}

\begin{document}

		\title{Non-Hermitian physics and master equations}

		\author{Federico Roccati}
		\affiliation{Universit\`a  degli Studi di Palermo, Dipartimento di Fisica e Chimica -- Emilio Segr$\grave{e}$, via Archirafi 36, I-90123 Palermo, Italy}
		
		\author{G.~Massimo Palma}
		\affiliation{Universit\`a  degli Studi di Palermo, Dipartimento di Fisica e Chimica -- Emilio Segr$\grave{e}$, via Archirafi 36, I-90123 Palermo, Italy}
		\affiliation{NEST, Istituto Nanoscienze-CNR, Piazza S. Silvestro 12, 56127 Pisa, Italy}
		
		\author{Fabio Bagarello}
		\affiliation{Universit\`a  degli Studi di Palermo, Dipartimento di Ingegneria, viale delle Scienze, I--90128  Palermo, Italy}
		
		\author{Francesco Ciccarello}
		\affiliation{Universit\`a  degli Studi di Palermo, Dipartimento di Fisica e Chimica -- Emilio Segr$\grave{e}$, via Archirafi 36, I-90123 Palermo, Italy}
		\affiliation{NEST, Istituto Nanoscienze-CNR, Piazza S. Silvestro 12, 56127 Pisa, Italy}
	
		\date{\today}
	
		\begin{abstract}

		A longstanding tool to characterize the evolution of open Markovian quantum systems is the GKSL (Gorini-Kossakowski-Sudarshan-Lindblad) 
		master equation. However, in some cases, open quantum systems can be effectively described with \textit{non-Hermitian Hamiltonians}, which have attracted great interest in the last twenty years due to a number of unconventional properties, such as the appearance of \textit{exceptional points}. 
		Here,  we present a short review of these two different approaches aiming in particular to highlight their relation and illustrate  
		different ways of  
		connecting non-Hermitian Hamiltonian
		to a GKSL master equation for the full density matrix.

		\end{abstract}
	
		\maketitle

		\section{Introduction}

		One of the postulates of quantum mechanics is that physical observables are described by self-adjoint\footnote{In most physical contexts one usually uses the term \textit{Hermitian}, having in mind the matrix representation of the operator. As it will not make any substantial difference, we will loosely use the terms \textit{self-adjoint} and \textit{Hermitian} interchangeably, despite these two concepts are not mathematically equivalent. } operators acting on the state space, i.e.~a Hilbert space~\cite{cohen2019quantum}. The reason is at least twofold: 
		\begin{itemize}
			\item[$(i)$] measurements outcomes, i.e.~the eigenvalues of an observable, must be real numbers and Hermiticity is a sufficient condition:
			\begin{equation}
			a\braket{\psi}=\braket{\psi}{\mathcal A\psi}=\braket{\mathcal A\psi}{\psi}=a^*\braket{\psi}
			\end{equation}
			where $\mathcal A=\mathcal A^\dagger$ is an observable, satisfying the eigenequation $\mathcal A\ket{ \psi} = a \ket{ \psi}$
			\item[$(ii)$] the eigentates corresponding to different measurement outcomes are distinguishable, i.e.~orthogonal, and form a complete set:
			\begin{eqnarray}
			a_2\braket{\psi_1}{\psi_2}
			& = & 
			\braket{\psi_1}{\mathcal A\psi_2} \nonumber\\
			& = & 
			\braket{\mathcal A\psi_1}{\psi_2}=a_1\braket{\psi_1}{\psi_2}\nonumber\\
			& \Rightarrow & 
			\braket{\psi_1}{\psi_2}=0
			\end{eqnarray}
			and
			\begin{equation}
			\mathbb{1}=\sum_n \frac{\dyad{\psi_n}}{\braket{\psi_n}}
			\end{equation}
			where $a_1\neq a_2$ and $\mathcal A\ket{ \psi_n} = a_n \ket{ \psi_n}$. 
		\end{itemize}

		
		When applied to the Hamiltonian operator $H$, the requirement of Hermiticity combined with the Schr\"odinger equation imply that
		the \textit{full} evolution must be unitary. This, in particular, results in the conservation of the wavefunction norm~\cite{cohen2019quantum}:
		\begin{equation}\label{consNorm}
		|\!\braket{\psi_t}\!|^2
		=
		|\!\bra{\psi_0}U^\dagger U\ket{\psi_0}\!|^2
		=
		|\!\braket{\psi_0}\!|^2 
		\end{equation}
		where $\ket{ \psi_0}$ is the initial state and $U=e^{-iHt}$ is the evolution operator for a time independent Hamiltonian $H$ ($\hbar=1$ throughout). 
		Therefore, \textit{Hermiticity} (of the Hamiltonian at least) appears a key assumption to describe \textit{closed} quantum systems. 
		
		However, no realistic physical system can be isolated from its external environment. This is even more compelling for quantum systems, as the act of measurement itself (i.e.~the interaction with an element of the environment) projects the state of the system onto an (or superposition of) eigenstate(s) of the measured observable corresponding to the measurement outcome. Being a projection, the act of measurement cannot be a unitary operation and therefore does not conserve probability.
		It is thus natural to wonder whether, dropping the assumption of Hermiticity, it is possible to describe, at least \textit{effectively}, the dynamics of \textit{open} quantum systems~\cite{benderAJoP2003,benderRPP2007}.
		
		This question is 
		is a possible way of introducing
		the field of \textit{non-Hermitian (quantum) physics}, a research area that has attracted great attention per se, beyond the initial motivations~\cite{BergholtzRMP2021,KawabataPRX2019,AshidaAiP2020}.
		
		The topic of non-Hermitian physics can be tackled from two different perspectives: on the one hand, non-Hermiticity can be regarded as an \textit{effective} description of an open system, which nonetheless shows 
		interesting intrinsic features
		(e.g.~appearance of exceptional points~\cite{heissJPAMT2012,Mirieaar7709}, non-Hermitian skin effect~\cite{BergholtzRMP2021}, etc.), under the constraint that the full microscopic Hamiltonian is always Hermitian. On the other hand, a more mathematically-oriented research line, which  was triggered by the idea that Hermiticity could be replaced by a symmetry assumption (\PT~symmetry~\cite{benderPRL1998}), is  assuming that the \textit{full} Hamiltonian of the system is generally non-Hermitian
		and on this basis either constructing
		the formalism of \textit{bi-orthogonal} quantum mechanics~\cite{brody2013biorthogonal}, or trying to somehow redefine the Hilbert space  so that  the Hamiltonian becomes Hermitian\footnote{The concept of Hermiticity  relies on the chosen inner product in the Hilbert space, which is of course not unique.}~\cite{mostafazadeh2010conceptual}, or finally investigating the mathematical framework of non self-adjoint physically-inspired operators~\cite{bagbook}.
		
		In this short review we will focus on the first, more physical, approach. 
		Before discussing the relation between the GKSL master equation  approach and 
		non-Hermitian Hamiltonians for the description of open (Markovian) quantum dynamics (Sec.~3), we present in the next section the role of \PT~symmetry as this was a starting point of this literature and most importantly because 
		a variety of platforms (especially in optics)
		implementing \PT~symmetry  have been experimentally realized.
		We then explicitly compare in Sec.~4 the Linblad and  non-Hermitian approach in the case study of atomic decay. In Sec.~5 we show how non-Hermitian dynamics at the mean-field level naturally arise in the context of Gaussian systems. Finally, in Sec.~6 we brielfy discuss how dissipative dynamics, regardless of the description, can yield exotic behaviors which are not achievable in closed dynamics, such as non-reciprocal excitation transfer. We then draw our conclusions.
		
		As said in the above, a major focus here is linking non-Hermitian Hamiltonians with Markovian master equations. Extensive review papers fully dedicated to non-Hermitian physics recently appeared such as Refs.~\cite{BergholtzRMP2021,KawabataPRX2019,AshidaAiP2020}, to which we refer the interested reader.

		\section{Non-Hermiticity, \PT~symmetry and exceptional Points}

\begin{figure}
			\centering
			\includegraphics[width=0.8\columnwidth]{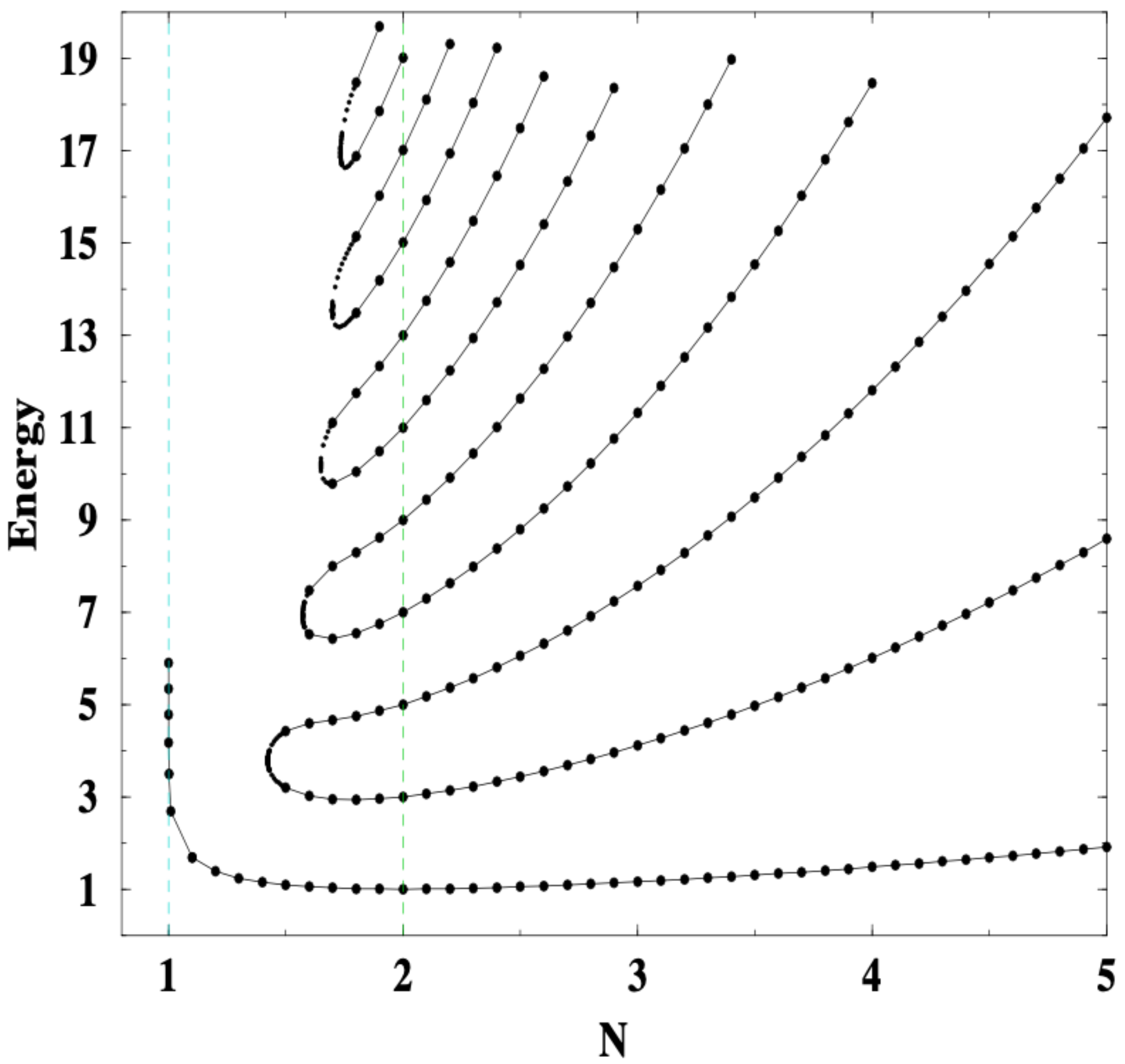}
			\caption{Spectrum of $H$ 
				in Eq.~\eqref{benderH} as a function of $N$. 
				If $N\geq2$ it is purely real, while  purely complex for $N\leq1$. 
				Reprinted by permission from the American Physical Society: Physical Review Letters~\protect\cite{benderPRL1998}, license number RNP/21/NOV/047109 (1998).
			} 
			\label{bender_spectrum}
		\end{figure}

		The recent history of non-Hermitian physics starts from the observation that some physically-inspired \textit{Hamiltonians} display real spectra of eigenvalues despite being non-Hermitian\footnote{It worth recalling that Hermiticity is only a sufficient condition for the spectrum to be real.}. The prototypical example is the non-Hermitian Hamiltonian initially proposed by Bender~\cite{benderPRL1998}
		\begin{equation}\label{benderH}
		H=p^2-(ix)^N.
		\end{equation} 
		This \textit{Hamiltonian} has a discrete and positive spectrum for $N\geq 2$, see Fig.~\ref{bender_spectrum}, and for $N>2$ is not self-adjoint\footnote{Except for even $N$.}. This property, first supported numerically and then proved mathematically~\cite{giordanelli2013real}, suggested  the idea  
		that occurrence of a real spectrum is related to the invariance
		of $H$ in Eq.~\eqref{benderH} under both \textit{parity} $\mathcal{P}$ and \textit{time reversal} $\mathcal{T}$, as represented by the equation $[H,\mathcal{PT}]=0$. 
		
		Operators satisfying this property are called \PT-symmetric. Note that the explicit form of $\mathcal{P}$ and $\mathcal{T}$ depends on the specific representation: e.g., for a continuous case as in Eq.~\eqref{benderH} they are defined as $\mathcal{P}: \{p,x\}\rightarrow\{-p,-x\}$ and $\mathcal{T}: i\rightarrow-i$. 
		
		Let us observe that, contrarily to what one is usually accustomed to in (Hermitian) quantum mechanics, the commutation relation $[H,\mathcal{PT}]=0$ does not imply that $H$ and \PT~have a common basis of eigenstates. Indeed this holds true if both commuting operators are Hermitian which is not the case here since \PT~is anti-Hermitian~\cite{cohen2019quantum}. 
		
		This points to the existence of two distinct \PT-symmetric phases: an \textit{unbroken} phase when $H$ and \PT~do possess a common basis of eigenstates, and a \textit{broken} \PT-symmetric phase when they do not~\cite{el-ganainyNP2018}.
		
		A typical \textit{discrete}  realization of \PT-symmetric Hamiltonians are \textit{gain-loss} systems, which we will discuss later in more details, especially because it is the scenario where \PT-symmetry breaking (from unbroken to broken) through the so called \textit{exceptional point} has been observed in classical optics~\cite{ruterNP2010}.
		
		\subsection{Gain-loss systems}
		
\begin{figure}
			\centering
			\includegraphics[width=\columnwidth]{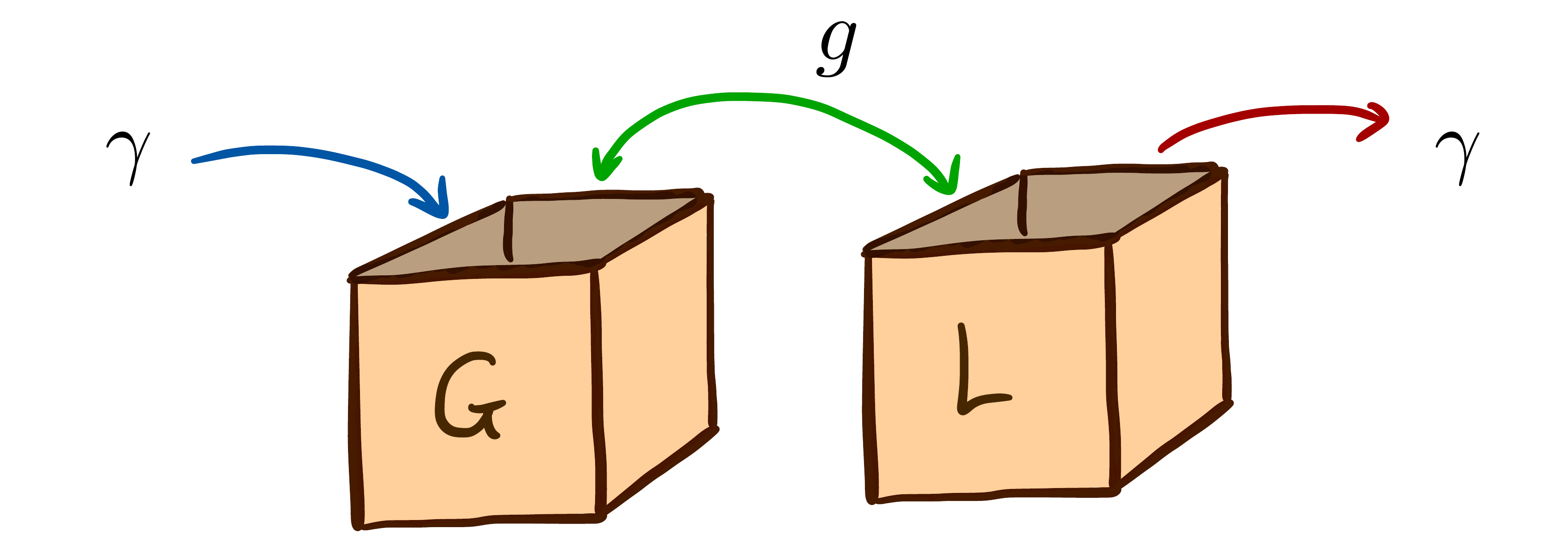}
			\caption{Sketch of a general gain-loss system
				.} 
			\label{gain_loss_box}
		\end{figure}

		We  report next  a very general (apparently unrelated) discussion based on~\cite{bender2019pt} in order to pedagogically introduce the topic. Consider two boxes whose contents (e.g., energy, matter, etc.)~are labeled by continuous (generally complex) variables $G$ and $L$. Suppose these two boxes are exchanging their content at rate $g$ and that the $G$ box is continuously filled (\textit{gain}) from the outside with rate $\gamma$, while  $L$  is leaking (\textit{loss}) into the environment at the same rate $\gamma$, see Fig.~\ref{gain_loss_box}. 
		
		As gain and loss rate are balanced (same rates), this ideal system is indeed \PT-symmetric: exchanging $G$ and $L$ (parity) and reversing the arrows in Fig.~\ref{gain_loss_box} (time reversal) 
		leaves the entire system invariant.
		
		If we want to quantitatively describe the dynamics of such a system, one way is to write the dynamical equations for $G$ and $L$: $\dot G = \gamma G -ig L$ and $\dot L = -\gamma L -ig G$ (the $\pi/2$ phase in the coupling is not necessary, but we keep it for the sake of  argument). These equations can be cast  into a \textit{Schr\"odinger-like} equation $i\dot\psi=\mathcal H\psi$ where $\psi=(G,L)^T$ and the \textit{Hamiltonian} reads 
		\begin{equation}\label{PT2by2}
		\mathcal H=\left(
		\begin{matrix} 
		i\gamma & g \\
		g & -i\gamma 
		\end{matrix} 
		\right).
		\end{equation}
		\PT~symmetry is then more manifest as 
		\begin{equation}
		\mathcal P =\left(\begin{matrix} 0 & 1 \\1 & 0 
		\end{matrix} \right)
		\end{equation}
		and $\mathcal T$ is complex conjugation, hence $[\mathcal{H},\mathcal{PT}]=0$. 
		
		The vector $\psi$ is not a state of a quantum system, nevertheless the analogy with the Schr\"odinger equation is not fully out of context. Indeed, as for any system of coupled differential equations, the solution can be written\footnote{Provided that $\mathcal H$  is diagonalizable} as $\psi_t=\sum_{j=\pm} c_j\varphi_je^{-i\lambda_jt}$, where $\varphi_{j}$'s are $\mathcal H$'s eigenvectors corresponding to the eigenvalues $\lambda_{j}$'s, and $c_{j}$'s are coefficients.
		
		As for the Bender's Hamiltonian in Eq.~\eqref{benderH}, one can either suppose that a Schr\"odinger equation $i\dot\psi=\mathcal H\psi$ with $\mathcal H$ as in Eq.~\eqref{PT2by2} makes sense in some suitably defined Hilbert space, or that $\mathcal H$ is an \textit{effective} Hamiltonian.
		

		Let us now examine the eigenvectors and eigenvalues of $\mathcal H$. 
		Its eigenvalues are $\lambda_\pm=\pm\sqrt{g^2-\gamma^2}$ with corresponding eigenvectors
		\begin{equation}
		\varphi_\pm 
		=
		\left(
		\begin{matrix} 
		i\gamma+\lambda_\pm  \\
		g
		\end{matrix} 
		\right). 
		\end{equation}
		A few observations are in order: 
		\begin{itemize}
			\item we can consider the parameter $\gamma$ 
			as the degree of non-Hermiticity
			since $H$ is Hermitian
			if and only if $\gamma=0$, 
			\item for any non-zero $\gamma$ the eigenstates are not orthogonal\footnote{With respect to the standard inner product in $\mathbb{C}^2$.},
			\item  for $\gamma<g$ ($\gamma>g$) eigenvalues are real (imaginary) and \PT~symmetry is unbroken (broken). Indeed if $\gamma<g$ the eigenstates $\varphi_\pm $ are eigenstates of both $\mathcal H$ and \PT, while if $\gamma>g$ they are not.
		\end{itemize}

		
		\subsection{Exceptional points}
		
		The critical value $\gamma=g$ found previously deserves special attention. At this point, not only the eigenvalues are degenerate (as it can occur for Hermitian operators), but even the 
		eigenvectors are  {\it coincident} (``coalesce'').
		
		This point of non-Hermitian degeneracy is called \textit{exceptional point} (EP): at an EP of order $n$ (in the case of Eq.~\eqref{PT2by2} we have $n=2$), $n$  eigenvalues \textit{together with the corresponding eigenstates} coalesce~\cite{el-ganainyNP2018}. 
		
		
		\begin{center}
			\begin{figure}[t]
				\centering
				\includegraphics[width=\columnwidth]{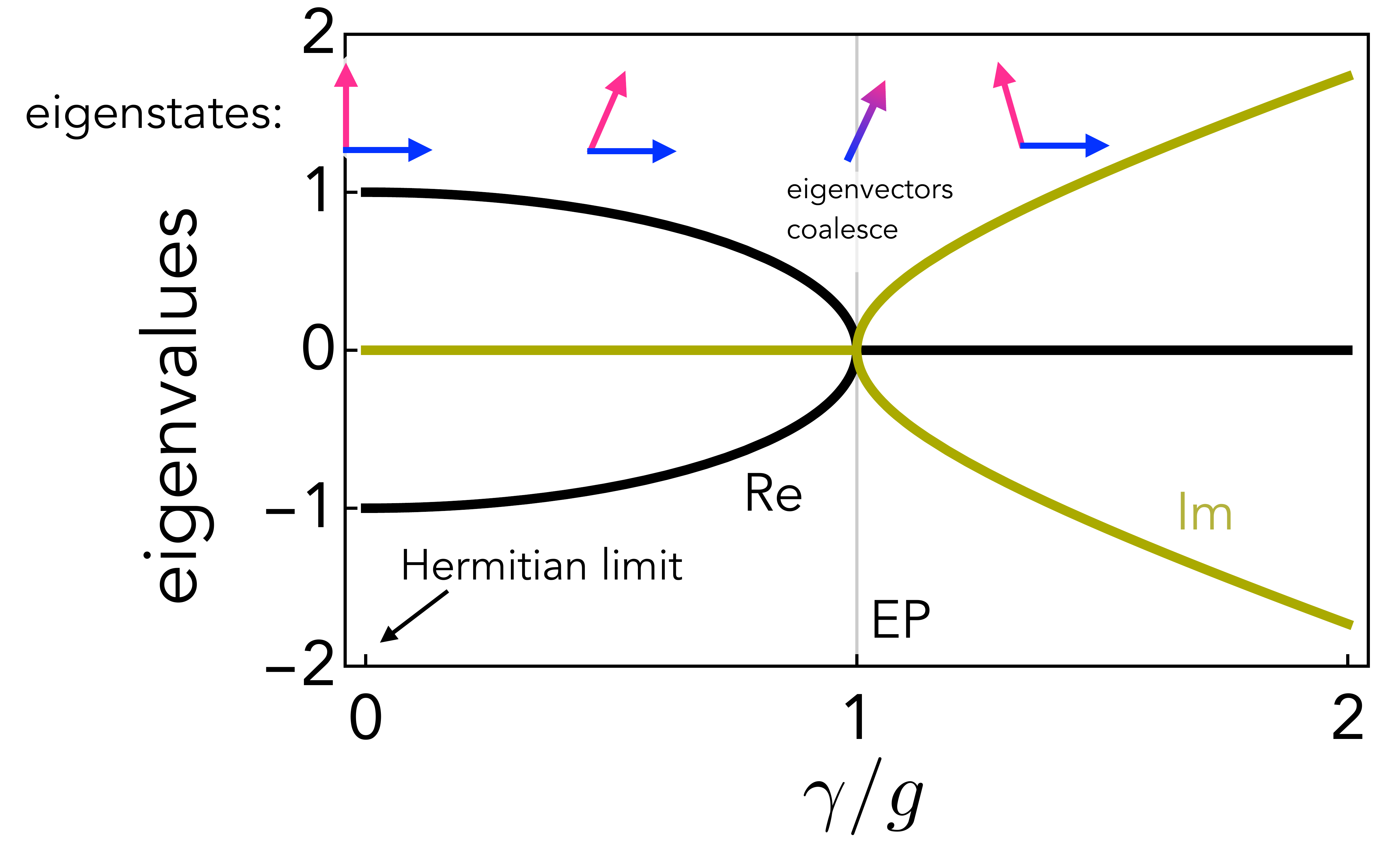}
				\caption{Spectrum of $\mathcal H$ in Eq.~\eqref{PT2by2} as a function of $\gamma/g$. Eigenvalues are real (imaginary) in \PT~unbroken (broken) phase $\gamma<g$ ($\gamma>g$), and coincide at EP. Eigenvectors are pictorially represented so as to highlight their orthogonality in the Hermitian limit and coalescence at the EP.} \label{EPgeneral}
			\end{figure}
		\end{center}
		
		Most remarkably, this entails that at an EP 
		the Hamiltonian becomes defective (diagonalizability is lost), that is its eigenstates do \textit{not} span the entire Hilbert space (the identity is not resolved).

		Exceptional points and the corresponding \PT~symmetry breaking transition 
		received great attention
		in recent years, and one of the main current interests in this field is the potential exploitation of critical behaviors near EPs in  quantum systems. For instance: enhancing mode splitting between counter-propagating whispering gallery modes in nanophotonics~\cite{Mirieaar7709}, EP-based sensors~\cite{WiersigPR2020}, and critical behavior of quantum correlations near EPs~\cite{roccati2021quantum}. 
		
		We want to stress that EPs are \textit{not} a prerogative of \PT~symmetry as they can appear for general non-Hermitian operators. A simple example is the case that our two boxes are not subject to gain or loss, yet the  coupling is non-reciprocal:
		\begin{equation}\label{nonR}
		\mathcal H'=\left(
		\begin{matrix} 
		0 & g_1 \\
		g_2 & 0 
		\end{matrix} 
		\right).
		\end{equation}
		with $g_2\neq g_1^*$.
		At $g_2=0$, the  eigenvalues and eigenvectors coalesce despite  $[\mathcal{H}',\mathcal{PT}]\neq0$.
		
		\subsection{An optical implementation of \PT~symmetry}
		
		One of the first implementations of \PT~symmetry was realized in classical optics~\cite{ruterNP2010}, as briefly described next. The starting observation is  the paraxial equation of diffraction
		\begin{equation}\label{PEoD}
		i\frac{\partial E}{\partial z}+\frac{1}{2k} \frac{\partial^2 E}{\partial x^2} + k_0 [n_R(x)+in_I(x)]E =0,
		\end{equation}
		describing the propagation along the $z$ direction of the electric field envelope $E$ of an optical beam. 
		
		This 
		can be regarded as an effective Schr\"odinger equation with a complex potential. Here $k_0$ ($k$) is the wave number in the vacuum (substrate), the $x$ direction is transversal to $z$ and  $n_{R/I}(x)$ are the real/imaginary part of the refracting index, playing the role of an optical potential. 
		
		By judiciously setting them so that $n_R(x)$ is even and $n_I(x)$ is odd, Eq.~\eqref{PEoD} has the form of a Schr\"odinger equation with a \PT-symmetric potential, in the spirit of the Bender's Hamiltonian~\eqref{benderH}. 
		
		Furthermore, under these assumptions, by using the coupled-mode approach~\cite{ruterNP2010}, the propagation of the electric field envelopes $E_1$ and $E_2$  in two coupled waveguides, the first being optically pumped (gain), the second experiencing the same amount of loss, is given by
		\begin{equation}\label{coupledmode}
		i\frac{d}{dz}
		\left(
		\begin{matrix} 
		E_1 \\
		E_2 
		\end{matrix} 
		\right)
		=
		\left(
		\begin{matrix} 
		i\gamma & g \\
		g & -i\gamma 
		\end{matrix} 
		\right)
		\left(
		\begin{matrix} 
		E_1 \\
		E_2 
		\end{matrix} 
		\right).
		\end{equation}
		
\begin{center}
			\begin{figure}[t]
				\includegraphics[width=0.9\columnwidth]{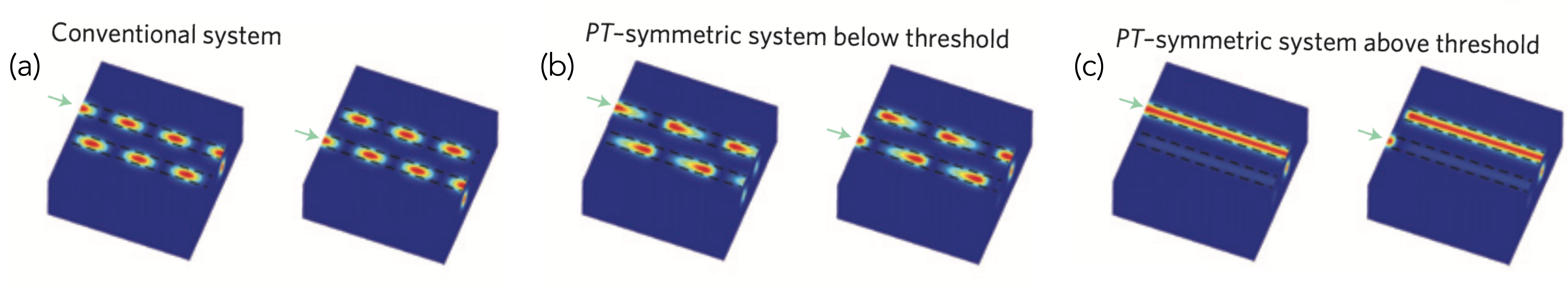}
				\caption{Propagation of light in coupled waveguides following Eq.~\eqref{coupledmode}. Left (Right): input in waveguide 1 (2). (a): Hermitian limit $\gamma =0$. (b): Unbroken \PT~symmetry $\gamma < g$. (c): Broken \PT~symmetry $\gamma > g$.
					Adapted by permission from Springer Nature Customer Service Centre GmbH: Springer Nature, Nature Physics~\protect\cite{ruterNP2010}, license number 5195341279265 (2010).
				} \label{couplemode}
			\end{figure}
		\end{center}

		This implements a gain-loss system, Eq.~\eqref{PT2by2} with all the properties outlined earlier. 
		The \PT-symmetric transition at the EP has been experimentally observed
		as light propagation strongly depends on the  \PT~phase, unbroken or broken.
		In the former, the optical wave propagates jumping back and forth between the waveguides somewhat in a Hermitian-like fashion. In the latter, only light injected in the pumped channel survives irrespectively of the initial input, see Fig.~\ref{couplemode}. 
		

		\onecolumngrid
		\twocolumngrid
		
		\subsection{Passive\,-\,\PT~symmetry}
		
		The  \textit{Hamiltonian} of a balanced gain-loss system is  
		probably the most typical way of introducing 
		\PT~symmetry breaking at EPs. 
		However, 
		implementing such a system can raise some issues.
		On the one hand, a  gain like the one previously considered is an approximation holding 
		up to not too long times
		(to avoid insurgence of non-linearities)~\cite{PurkayasthaPRR2020}. On the other hand it is nonsensical when working ab initio with nonlinear systems (such as a two-level atom).

		However, most of the above phenomena
		(appearance of EPs, different dynamical behaviors below and above EP, etc.) still hold true in fully dissipative systems (no gain whatsoever), 
		for instance when one introduces {\it non-uniform} losses.
		
		As an instance, consider again a pair of coupled 
		waveguides under the coupled-mode approach. If they experience a different amount of loss 
		then the \textit{Hamiltonian} reads
		\begin{equation}\label{passPT2by2}
		\mathcal H
		=\left(
		\begin{matrix} 
		-i\gamma_1 & g \\
		g & -i\gamma_2 
		\end{matrix} 
		\right).
		\end{equation}
		The corresponding spectrum and eigenstates are
		\begin{equation}
		\lambda_\pm=-i\Gamma_+\pm\sqrt{g^2-\Gamma_-^2}\,\,,
		\quad
		\varphi_\pm 
		=
		\left(
		\begin{matrix} 
		\lambda_\pm+i\gamma_2  \\
		g
		\end{matrix} 
		\right)
		\end{equation}
		where $2\Gamma_\pm=\gamma_2\pm\gamma_1$. 
		
		Despite losing \PT~symmetry ($[\mathcal H,\mathcal{PT}]\neq0$), there still is an EP at $\Gamma_-=g$  separating two dynamical dissipative phases, an \textit{underdamped} one ($\Gamma_-<g$) and an \textit{overdamped} one ($\Gamma_->g$). 
		
		The connection with \PT~symmetry is  manifest once we notice that by making a complex global shift on the balanced gain-loss Hamintonian $\mathcal H$ in Eq.~\eqref{PT2by2}
		\begin{equation}
		\mathcal H\rightarrow\mathcal H-i\gamma\mathbb{1}
		=
		\left(
		\begin{matrix} 
		0 & g \\
		g & -2i\gamma 
		\end{matrix} 
		\right)
		\end{equation}
		one gets a passive-\PT-symmetric Hamiltonian with only one local loss and no gain. 
		This operation shifts the spectrum to the lower half of complex plane, without altering the eigenvectors. Therefore, the key point of these so called  \textit{passive}-\PT-symmetric systems~\cite{Ornigotti_2014} is that we have ${\rm Im}( \lambda^p_\pm)\leq0$ avoiding any amplification process, but still inheriting  features of \PT-symmetric systems~\cite{JoglekarPRes2018}.

		\section{Open Markovian quantum dynamics}

		The evolution of closed quantum systems is governed by the Schr\"odinger equation
		\begin{equation}\label{SEkets}
		i\frac{\rm  d}{{\rm d}t}\ket{ \Psi_t}= H \ket{ \Psi_t}\,,
		\end{equation} 
		where $\ket{ \Psi_t}$ is the state of the whole system at time $t$ living in the total Hilbert space $\mathscr H$, and $H$ is the (Hermitian) Hamiltonian of the full system acting on $\mathscr H$. 
		
		By \textit{open} quantum system, we mean one that is interacting with an \textit{environment} so that the full dynamics of system  and environment  is unitary~\cite{breuer2007,rivas2012open,ciccarello2021quantum}. This separation of system $S$ and environment $E$ corresponds to a bipartition of the Hibert space, $\mathscr H=\mathscr H_{S}\otimes\mathscr H_{E}$, 
		with $\mathscr H_S$ ($\mathscr H_E$) the system's (environment's) Hilbert space.
		
		When dealing with composite systems, the description of  physical states through \textit{kets} (or wave functions) as in Eq.~\eqref{SEkets}  is 
		no longer complete. Indeed,
		just to make an example, if  $\psi(x_1,x_2)$ is the wave function of two  particles, there is 
		no way to infer from it a well-defined wavefunction
		of the first particle only~\cite{hall2013} (unless the joint state is factorized). 
		
		The most general description of physical states is indeed
		through \textit{density operators} instead of kets, allowing 
		in particular an appropriate
		description of \textit{subsystems}~\cite{landau1965}. More concretely, if $\ket{ \Psi_t}$ is the state of $S+E$ at time $t$, then the corresponding density operator is given by  $\rho_t=\dyad{\Psi_t}$ and density operators representing $S$ and $E$ are obtained by tracing out $E$ and $S$, respectively as $\rho_{{S(E)}}=\Tr_{{E(S)}}\rho$.

		In the language of density operators, the Schr\"odinger equation turns into the \textit{von Neumann equation}
		\begin{equation}\label{VNeq}
		\dot\rho_t=-i[H,\rho_t].
		\end{equation} 
		Although containing the same amount of information as Eq.~\eqref{SEkets} for the full $S+E$ dynamics, 
		the density-matrix formalism allows to
		(at least formally) write the reduced dynamical equation for the system (or environment) only, that is
		\begin{equation}\label{VNeqsystem}
		\dot\rho_{S}=-i\Tr\!_{E}[H,\rho_t].
		\end{equation}
		At this level, this equation is not closed in $\rho_S$ (i.e., it is not a so called master equation). 
		However, under the assumptions that 
		\begin{itemize}
			\item[$(i)$]  $\rho_t\approx\rho_S(t)\otimes\rho_E$, i.e.~the \textit{Born approximation},
			\item[$(ii)$] the time scale  over which the
			state of the system varies appreciably is large compared to the time scale  over which the environment correlation functions decay (\textit{Markov approximation}),
			\item[$(iii)$] the \textit{rotating-wave approximation} for system-environment interaction holds,
		\end{itemize}
		the   
		master equation for $S$
		can be written as~\cite{breuer2007}
		\begin{equation}\label{lindblad}
		\dot\rho_{S} = -i[H_{S},\rho_{S}] + \mathscr D(\rho_{S})
		\end{equation}
		where $H_{S}$ is the free Hamiltonian of the system
		and the \textit{dissipator} reads
		\begin{equation}\label{dissipator}
		\mathscr D(\rho_{S}) = \sum_{i } \Gamma_{i} \left(\hat L_i \rho_{S} \hat L_i^\dagger - \frac{1}{2} \{ \hat L_i^\dagger \hat L_i,\rho_{S} \} \right)
		\end{equation}
		where $\{\hat L_i\}$ are the so called \textit{jump} operators acting on $\mathscr H_{S}$ and the rates $\{\Gamma_{i}\}$ are positive. 
		
		Equation~\eqref{lindblad} is called GKSL (Gorini -Kossakowski-Sudarshan-Lindblad)  master equation and describes the dynamics of an open quantum system interacting with an environment, under the previously discussed assumptions. 
		A large variety of physical systems are well-described  by such equation.
		
		In the following 
		will discuss some paradigmatic instances showing the connection between the previously discussed non-Hermitian physics with the GKSL master equation. 
		

		\section{Atom decay: GKSL master equation versus non-Hermitian Hamiltonian}
		
		
		A minimal model for atomic decay is that of a two-level system (the atom) interacting with a zero temperature thermal reservoir~\cite{breuer2007}. The two-level approximation works in every situation where only the transition between two levels is significant and all others can be neglected. 
		
		We call these two levels $\ket{g}$ and $\ket{e}$ standing for \textit{ground} and \textit{excited} state, respectively,
		whose respective energies are
		$\omega_g=0$ and $\omega_e$. The free  Hamiltonian of the atom thus reads $H_{S}=\omega_e\dyad{e}$. In order to study atomic decay we will consider $\ket{e}$ as the initial state of the system. 
		
		\subsection{Decay of a two-level system: master equation description}
		
		First, we describe atomic decay
		through the GKSL master equation as in~\cite{breuer2007,CTap1992}.
		Coupling the atom to the quantized radiation field through 
		the electric-dipole approximation and rotating-wave approximation,
		and assuming the field to be in a zero temperature thermal state (vacuum) master equation~\eqref{lindblad} in this case reads
		\begin{equation}\label{2levLind}
		\dot\rho = -i[H_{S},\rho] + \Gamma \left( \sigma_- \rho \sigma_+ - \frac{1}{2} \{ \sigma_+\sigma_-,\rho \} \right)
		\end{equation}
		where $\rho$ is the state of the two-level system, $\sigma_-=\dyad{g}{e}$ is the jump operator, $\sigma_+=\sigma_-^\dagger$ and $\Gamma>0$. 
		
		In the $\{ \ket{e},\ket{g} \}$ basis the state is represented by
		\begin{equation}\label{state2lev}
		\rho=
		\left(
		\begin{matrix} 
		\rho_{ee} & \rho_{eg} \\
		\rho_{ge} & \rho_{gg} 
		\end{matrix} 
		\right),
		\end{equation}
		where $\rho_{ij}=\bra{i}\rho\ket{j}$, $\rho_{ee(gg)}$ is the \textit{population} of the excited (ground) state and $\rho_{eg,ge}$ are the \textit{coherences}. The solution of Eq.~\eqref{2levLind} is then
		\begin{equation}\label{state2levSol}
		\rho_t=
		\left(
		\begin{matrix} 
		e^{-\Gamma t} & 0 \\
		0 & 1-e^{-\Gamma t} 
		\end{matrix} 
		\right)
		\end{equation}
		and  correctly captures the fact that, starting from the excited state, population is irreversibly transferred to the ground state: $\rho_0=\dyad{e}\rightarrow\rho_\infty=\dyad{g}$. Accordingly, the probability of finding the system in the excited (ground) state exponentially approaches 0 (1), see Fig.~\ref{LindbladVSnh}.
		
		\subsection{Decay of a two-level system: non-Hermitian description}
		
		A second more phenomenological approach to the instability of a state, 
		which is typical in non-Hermitian
		physics  
		(see e.g.~Ref.~\cite{cohen2019quantum}) can be formulated as follows.

		One can  solve the  Schr\"odinger equation corresponding to the free atomic Hamiltonian $H_S=\omega_e |e\rangle\langle e|$ getting $\ket{ \psi_t}=e^{-i\omega_e t}\ket{e}$ and then introduce 
		an {\it ad hoc} complex shift through the replacement
		$\omega_e\rightarrow\omega_e-i\Gamma/2$ so that $p_e=|\!\braket{\psi_t}{e}\!|^2=e^{-\Gamma t}$. This shift can of course be made at the Hamiltonian level $H_{S}\rightarrow (\omega_e-i\Gamma/2)\dyad{e}$, so as to introduce a non-Hermitian Hamiltonian from the beginning. 
		
\begin{center}
			\begin{figure}[t]
				\includegraphics[width=\columnwidth]{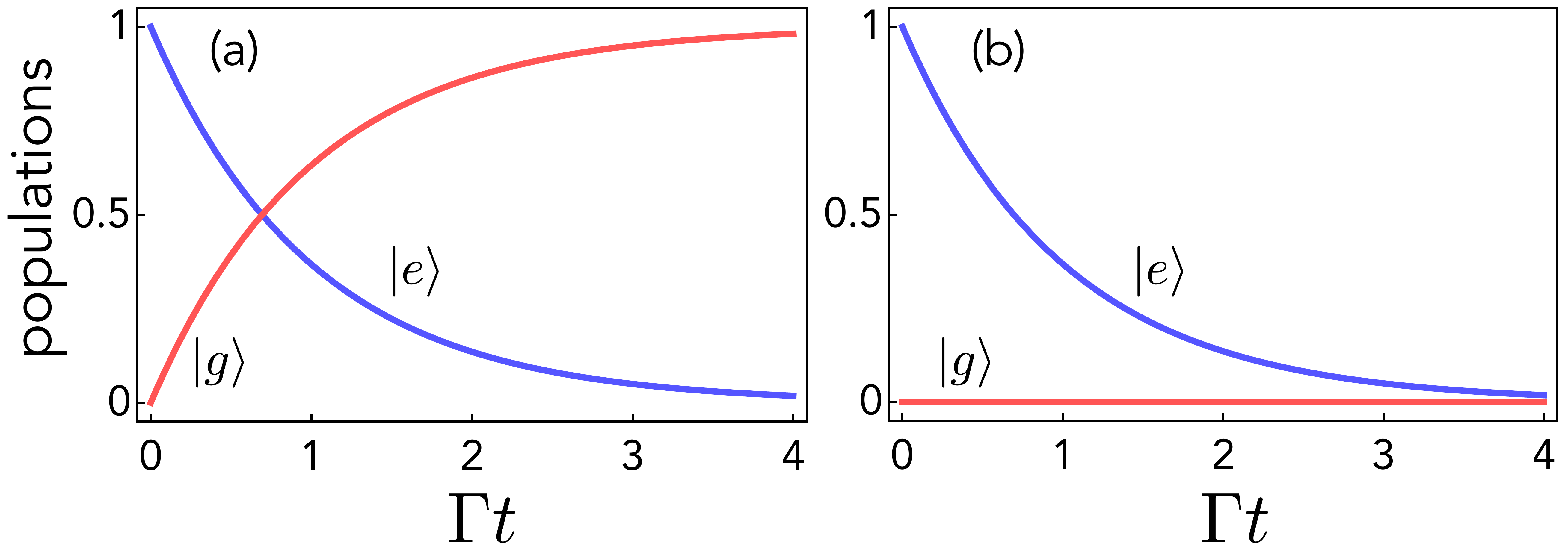}
				\caption{Populations of excited and ground state of a two-level system according to the  GKSL master equation (a) and  the phenomenological non-Hermitian description (b).} \label{LindbladVSnh}
			\end{figure}
		\end{center}

		Such approach correctly reproduces the probability $p_e$ of finding the two-level system in the excited state, however note that $p_g=|\!\braket{\psi_t}{g}\!|^2=0$. This  highlights how the non-Hermitian description is only an  \textit{effective} one, compared to the GKSL master equation. The latter indeed is a completely positive trace-preserving dynamics, while the former is not, as witnessed in particular by the non-conservation of the norm $|\!\braket{\psi_t}\!|^2=e^{-\Gamma t}$.
		Note that the GKSL master equation~\eqref{2levLind} correctly predicts that when the excited-state probability decays the ground-state population grows accordingly. The latter effect is instead absent in the non-Hermitian Hamiltonian description which only predicts excited-state probability, without  ground-state population growth, see Fig.~\ref{LindbladVSnh}.
		
		
		\subsection{Connection between master equation and non-Hermitian approach}
		
		The general  connection between the two approaches can be made more transparent by rearranging terms in the Linblad master equation~\eqref{lindblad} (subscript $S$ is omitted) as
		\begin{eqnarray}\label{lindbladRearrange}
		\dot\rho
		& = & -i[H,\rho] + \sum_{i } \Gamma_{i} \left(\hat L_i \rho \hat L_i^\dagger - \frac{1}{2} \{ \hat L_i^\dagger \hat L_i,\rho\} \right)
		\nonumber \\
		& = & -i\left( H_{\rm eff}\,\rho -\rho H_{\rm eff}^\dagger \right) + \sum_{i} \Gamma_{i} \hat L_i \rho \hat L_i^\dagger \,.
		\end{eqnarray}
		Here, 
		\begin{equation}
		H_{\rm eff} = H - \frac{i}{2} \sum_{i } \Gamma_{i} \hat L_i^\dagger \hat L_i
		\end{equation}
		is the effective non-Hermitian Hamiltonian. Dropping \textit{ quantum jumps} [last sum in Eq.~\eqref{lindbladRearrange}] one gets 
		\begin{equation}\label{NHlindblad}
		\dot\rho = -i\left( H_{\rm eff}\rho -\rho H_{\rm eff}^\dagger \right)
		\end{equation}
		which is  equivalent to working with a non-Hermitian Hamiltonian from the beginning~\cite{AshidaAiP2020}. 
		Observe that, being $H_{\rm eff} $  non-Hermitian, Eq.~\eqref{SEkets} implies $\frac{\rm  d}{{\rm d}t}\bra{ \Psi_t}= iH_{\rm eff} ^\dagger \bra{ \Psi_t}$ so that Eq.~\eqref{NHlindblad} follows.

		Of course discarding quantum jumps is as phenomenological as introducing a complex shift in the energy, and it can be regarded as a \textit{semiclassical limit} of the full quantum dynamics~\cite{mingantiPRA2019}. 
		Furthermore, the non-Hermitian Hamiltonian $ H_{\rm eff}$  emerges when dealing with quantum trajectories and post-selection~\cite{mingantiPRA2019}. 
		
		In the formalism of quantum trajectories, the state of the system is described by a stochastic wave function. For trajectories where no quantum jumps occur, the system evolves according to $ H_{\rm eff}$. On the other hand if a jump occurs, the state abruptly changes due to the term $\sum_{i } \Gamma_{i} L_i \rho L_i^\dagger$. 
		The GKSL  master equation~\eqref{lindblad} can be regarded as an average over infinitely many trajectories, i.e.~many experimental realizations. 
		
		Finally, 
		in order to illustrate the connection
		between the GKSL master equation and non-Hermitian Hamiltonian for the previously introduced decay of a two-level system we can rewrite Eq.~\eqref{2levLind} as 
		\begin{eqnarray}\label{2levLindnh}
		\dot\rho 
		& = & -i \left[ (\omega_e-i\Gamma/2)\dyad{e}\rho - \rho(\omega_e+i\Gamma/2)\dyad{e} \right] \nonumber\\
		&    &+ \Gamma \sigma_- \rho \sigma_+\,.
		\end{eqnarray}
		Except for the quantum jump term $\Gamma \sigma_- \rho \sigma_+$, this  is equivalent to working with the non-Hermitian Hamiltonian $(\omega_e-i\Gamma/2)\dyad{e}$.

		\section{Non-Hermitian mean-field dynamics from GKSL master equation}

		In this section we want to show an alternative method to derive non-Hermitian \textit{effective} Hamiltonians from a full Markovian master equation. 
		
		Consider a quantum system made of $N$ identical bosonic modes $\hat a_i$ ($[\hat a_i,\hat a_j^\dagger]=\delta_{ij}$) coherently exchanging excitations through the quadratic Hamiltonian
		\begin{equation}
		H=\sum_{i\neq j}(g_{ij}\hat a_i^\dagger \hat a_j+ {\rm H.c.} )
		\end{equation}
		where all of them are generally subject to local dissipation and incoherent pumping so that the full GKSL master equation reads
		\begin{equation}\label{quadraticlindbladian}
		\dot\rho= -i[H,\rho]+
		\sum_i \gamma_i \mathscr D[\hat a_i]\rho + \Gamma_i \mathscr D[\hat a_i^\dagger]\rho.
		\end{equation}
		where $\mathscr D[\hat A]\rho=\hat A\rho \hat A^\dagger -\{\hat A^\dagger \hat A,\rho\}/2$ and $\gamma_i,\Gamma_i>0$. We assumed  
		a rotating frame so as to eliminate the free Hamiltonian term $\omega_0 \sum_j\hat a_j^\dag\hat a_j$, with $\omega_0$ the frequency of each oscillator.
		
		
		We will restrict to the relevant class of \textit{Gaussian states}, i.e.~those states whose characteristic function is Gaussian~\cite{olivares2012quantum}. 
		These states are completely determined by the mean-field vector $\psi=(\alpha_1,\ldots,\alpha_N)^T$, with $\alpha_i=\expval{\hat a_i}$, and covariance matrix $\sigma_{ij} = \langle \hat A_i \hat   A_j+\hat A_j \hat A_i\rangle -2 \langle \hat A_i \rangle \langle \hat A_j \rangle$, where\footnote{Both mean-field vector and covariance matrix can be written in terms of position and momentum variables. 
			Here we stick to the representation with ladder operators.} $\hat A_i=(\hat a_1,\ldots, \hat a_N, \hat a_1^\dagger,\ldots, \hat a_N^\dagger)$~\cite{gardiner2004}. 
		
		For the GKSL master equation in Eq.~\eqref{quadraticlindbladian} the mean-field dynamics is given by  
		\begin{equation}
		i\dot\psi=\mathcal H\psi
		\end{equation}
		with 
		\begin{equation}\label{gainlossgeneral}
		\mathcal H = i(\mathcal G -\mathcal L) + \mathcal C 
		\end{equation}
		where $\mathcal G=$ diag$(\Gamma_1,\ldots,\Gamma_N)$, $\mathcal L=$ diag$(\gamma_1,\ldots,\gamma_N)$ and $\mathcal C_{ij}=g_{ij}$. 
		
		The mean-field dynamics is then  a Schr\"odinger-like equation with a non-Hermitian \textit{Hamiltonian} which, as previously discussed, despite having nothing genuinely quantum, can possess unconventional features like exceptional points or \PT~symmetry.
		
		A specific instance of such non-Hermitian mean-field dynamics coming from a GKSL master equation is  a gain-loss system~\cite{roccati2021quantum}. 
		This is a pair of quantum
		harmonic oscillators labeled by $G$ and $L$, whose joint state 
		evolves in time according to the GKSL master equation 
		\begin{eqnarray}\label{ME}
		\dot\rho  
		& = &  - i[g( \hat a_{L }^\dagger \hat a_{G}+{\rm H.c.}) ,\rho] \nonumber\\
		&  & +2\gamma_{L}\, \mathscr D[\hat a_L]\rho +2\gamma_{G} \,\mathscr D[\hat a_G^\dag]\rho 
		\end{eqnarray}
		which is a particular instance of Eq.~\eqref{quadraticlindbladian}.
		
		Besides the coupling Hamiltonian describing a coherent energy exchange at rate $g$ between the modes, the dissipators describe a local incoherent interaction    with a local environment: the one on $G$ pumps energy into the system with characteristic rate $\gamma_G$ (gain) while that on $L$ absorbs energy with rate $\gamma_L$ (loss). 
		
		This system can be implemented in a variety of ways \cite{el-ganainyNP2018}, including coupled waveguides \cite{ruterNP2010}, microcavities \cite{pengNP2014} and in  double-quantum-dot circuit QED setups~\cite{PurkayasthaPRR2020}.
		As for the general case of $N$ modes, from Eq.~\eqref{ME} it follows that the evolution of the mean-field vector $\psi=(\langle \hat{a}_L\rangle, \langle \hat{a}_G\rangle)^T$  is governed by the Schr\"odinger-like equation $i \dot{\psi} =  \mathcal H \psi$ with 
		\begin{equation}\label{evolMeanVal}
		\mathcal H=\left(
		\begin{matrix} 
		-i\gamma_{L } & g \\
		g & i\gamma_{G} 
		\end{matrix} 
		\right),
		\end{equation}
		which is exactly the non-Hermitian Hamiltonian in Eq.~\eqref{PT2by2} with generally different gain and loss rate, and a particular instance of Eq.~\eqref{gainlossgeneral}.
		If   gain  and loss are balanced ($\gamma_L=\gamma_G=\gamma$), $\cal H$ is \PT-symmetric.
		Equations analogous to~(\ref{ME}) for the full-quantum description of \PT-symmetric systems can be found also e.g.~in Refs.~\cite{PurkayasthaPRR2020,dastPRA2014,longhi2019quantum}.

		\section{Dissipation-induced non-reciprocity}

		Besides local dissipation/gain, non-Hermiticity can also come about as non-reciprocal
		%
		coupling between levels/modes (off-diagonal matrix elements), as we briefly discussed in Sec.2~\cite{BergholtzRMP2021,LieuPRB2018}, see Eq.~\eqref{nonR}. 
		
		A paradigmatic system in this respect is
		the \textit{Hatano-Nelson} model~\cite{HatanoPRL96}. It consists of a simple one dimensional tight-binding Hamiltonian 
		reading
		\begin{equation}\label{HatanoPBCS}
		H =
		\sum_{n} J(1+\delta)\dyad{n+1}{n} + J(1-\delta)\dyad{n}{n+1}
		\end{equation}
		where $J>0$, $-1\leq\delta\leq1$ and the lattice sites $\{\ket{n} \}$ can either represent levels, e.g.~in synthetic lattice models~\cite{WangSCIENCE2021}, or bosonic resonant modes (cavities) coupled to each other. Note that, for $\delta\neq 0$, $H_{n,n+1}\neq H_{n+1,n}^*$. This property in the present context is referred to as {\it non-reciprocal} couplings: the hopping rate from left to right differs from the one from right to left.
		
		This model can display peculiar non-Hermitian features besides exceptional points, such as high spectral sensitivity to boundary conditions accompanied by the so called \textit{non-Hermitian skin effect}~\cite{BergholtzRMP2021,RoccatiPRA2021,alvarez2018topological}, that is the accumulation of \textit{all} bulk eigenstates on a lattice edge under open boundary conditions.
		
		
		Despite being the minimal model where one could study peculiar non-Hermitian features, 
		implementing the Hatano-Nelson model  is non-trivial because of the non-reciprocal nature of the couplings~\cite{WangSCIENCE2021,LonghiSciRep2015}.
		
		In the following we will briefly outline the relationship between of non-reciprocity and the \textit{openness} of the quantum system 
		and the open dynamics of the system as described by the GKSL master equation.
		
		\subsection{Non-reciprocity via collective jump operators in a simple lattice}
		
		We will only show how to implement a single non-reciprocal coupling between two cavities (bosonic modes) as the generalization to an array of cavities is straightforward~\cite{MetelmannPRX2015}. 
		\begin{figure}[t]
			\centering
			\includegraphics[width=\columnwidth]{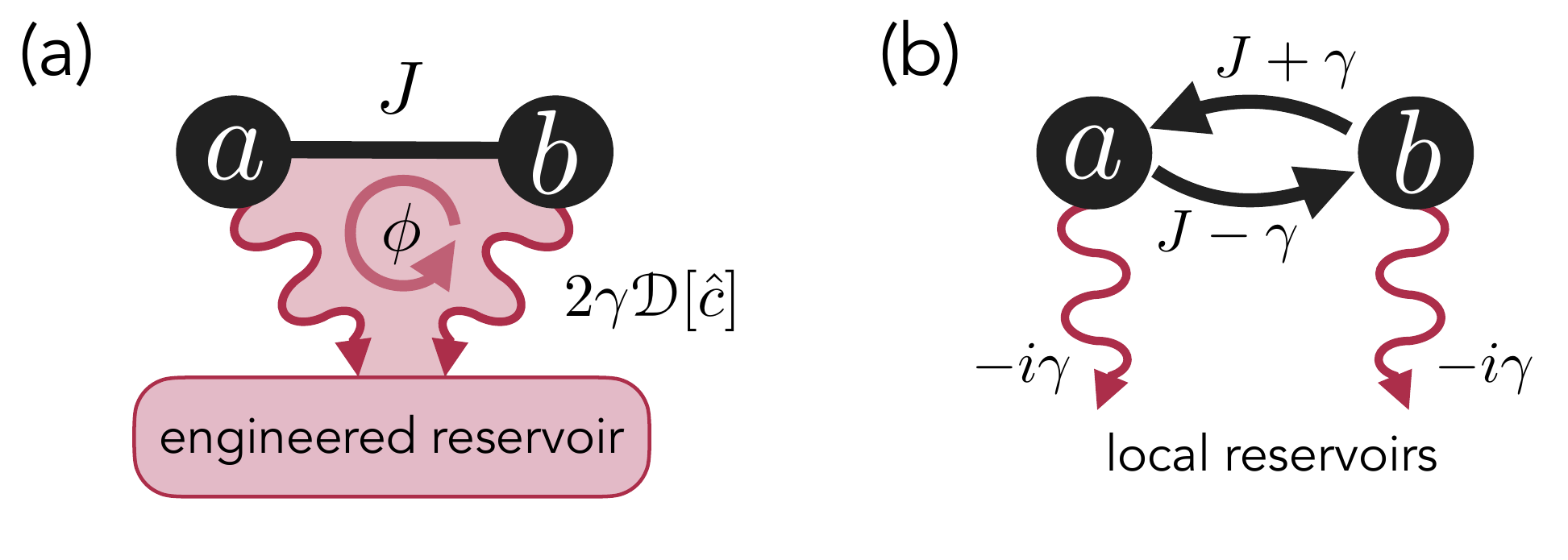}
			\caption{Non-reciprocity induced by engineered dissipation. (a) Two bosonic modes coupled to a common engineered bath, and to each other as given by the master equation in Eq.~\eqref{reseng}. (b) Same system (without considering quantum jumps) as described by the non-Hermitian Hamiltonian in Eq.~\eqref{NHhamcler} with $\phi=\pi/2$. Comparing (a) and (b) one can observe the connection between non-reciprocal couplings and a collective jump operator. } \label{NRclerk}
		\end{figure}
		
		Consider two modes $\hat a$ and $\hat b$ coherently interacting through the Hamiltonian $H=J(\hat a ^\dagger \hat b + \hat b ^\dagger \hat a )$ whose joint state evolves according to the master equation
		\begin{equation}\label{reseng}
		\dot\rho= -i[H,\rho]+2\gamma \mathscr D[\hat c]\rho .
		\end{equation}
		where 
		\begin{equation}\label{collJump}
		\hat c=\hat a + e^{i\phi} \hat b.
		\end{equation}
		This describes the interaction with a common bath through the  \textit{collective} (i.e., non-local) jump operator $\hat c$, cf.~Fig.~\ref{NRclerk}(a). 
		Replacing~\eqref{collJump} in Eq.~\eqref{reseng}, this can be arranged as
		\begin{equation}
		\dot\rho=-i\left( H_{\rm eff}\,\rho -\rho H_{\rm eff}^\dagger \right) 
		\,+\,{\rm quantum\,~jumps} 
		\end{equation}
		where the non-Hermitian Hamiltonian reads~\cite{MetelmannPRX2015}
		\begin{eqnarray}\label{NHhamcler}
		H_{\rm eff} 
		& = & 
		-i\gamma (\hat a ^\dagger \hat a + \hat b ^\dagger \hat b)+
		\left(J-i\gamma e^{i\phi} \right) \hat a ^\dagger \hat b \nonumber\\
		&  &
		+
		\left(J-i\gamma e^{-i\phi} \right) \hat b ^\dagger \hat a.
		\end{eqnarray}
		
		We observe how non-Hermiticity is not only due to local dissipation (diagonal terms), but generally also to non-reciprocity (off-diagonal terms). 
		The latter occurs by adjusting phase $\phi=\pi/2$,  so that we get a probability amplitude of hopping from $\hat b$ to $\hat a$ equal to $J+\gamma$ while the reverse process has a probability amplitude  $J-\gamma$, cf.~Fig.~\ref{NRclerk}(a).

		\subsection{Non-reciprocity via local jump operators in a structured lattice}
		
		The second approach we want to discuss takes a different perspective. In the previous one, the lattice structure (without dissipators) had a minor role in that non-reciprocal couplings in the open system arise specifically because of the non-local nature of  jump operators. 
		Despite the theoretical simplicity of the previous method, one could argue that the technically challenging task is to realize such a collective environment for every pair of modes in the lattice~\cite{MetelmannPRX2015}.
		
		An alternative complementary approach~\cite{ClerkArxiv2022} is asking whether, 
		provided that one engineers a possibly non-trivial
		lattice structure (complex or diagonal couplings), \textit{local} dissipation can yield some form of non-reciprocity. 

\begin{figure}[t]
			\centering
			\includegraphics[width=\columnwidth]{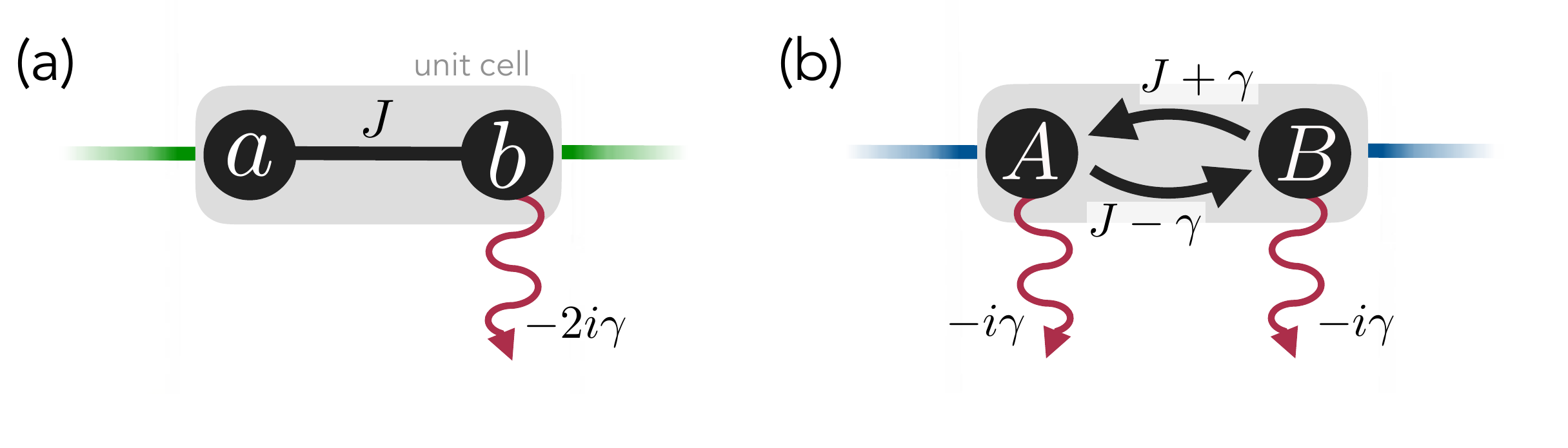}
			\caption{Non-reciprocity induced by non-uniform losses. (a) In grey the unit cell of a tight binding lattice made of bosonic modes is highlighted. The intracell coupling and dissipation is described through the non-Hermitian Hamiltonian in Eq.~\eqref{intercellHam}. The details on intercell couplings (green) are not relevant. Through the unitary transformation in~\eqref{intercellHamNewPic} on the entire lattice, yet local on the unit cell, lattice (a) transforms into (b). In the latter picture, besides local dissipation on \textit{all} lattice sites, we get non-reciprocal intracell couplings [intercell couplings (blue) are also modified by the transformation, but for the sake of argument we are concerned here on intracell features].} \label{NRcreutz}
		\end{figure}

		Consider the instance of a composite lattice, c.f.~Fig.~\ref{NRcreutz}, 
		with two cavities per unit cell
		labelled by $(\hat a_n,\hat b_n)$ so that the intracell Hamiltonian is 
		\begin{equation}\label{intercellHam}
		H_n
		=
		\left(
		\begin{matrix} 
		0 & J \\
		J & -2i\gamma 
		\end{matrix} 
		\right),
		\end{equation}
		representing coherent coupling between $\hat a_n$ and $\hat b_n$ and local dissipation on $b$-cavities, cf.~Fig.~\ref{NRcreutz}. 
		
		Non-reciprocity becomes manifest by changing picture. Indeed, by applying a unitary transformation, local on the unit cell, $(\hat a_n,\hat b_n)\rightarrow (\hat A_n,\hat B_n)$ so that $\hat a_n=\frac{1}{\sqrt{2}}(\hat A_n-i\hat B_n)$ and $\hat b_n=-\frac{i}{\sqrt{2}}(\hat A_n+i\hat B_n)$ the intracell Hamiltonian in this new picture becomes~\cite{BergholtzRMP2021}
		\begin{equation}\label{intercellHamNewPic}
		H_n \xrightarrow{U} H_n' 
		=
		\left(
		\begin{matrix} 
		-i\gamma & J+\gamma \\
		J-\gamma & -i\gamma 
		\end{matrix} 
		\right).
		\end{equation}
		Remarkably, local dissipation is now uniform but
		non-reciprocal couplings appear. 
		This argument shows how directional coupling can be achieved by judiciously engineering local dissipation in the photonic lattice. 
		
		Finally, we briefly mention that such a non-Hermitian photonic structure can yield exotic phenomena in a waveguide QED setup, as discussed in Ref.~\cite{roccatiArxiv2021}. Remarkably, by weakly coupling 
		emitters such as two-level atoms or resonators
		to an array of coupled cavities with patterned local dissipation, it turns out that the non-Hermitian lattice
		can mediate second-order (dissipative) interactions between the emitters with unique properties which are unusual if not unachievable in Hermitian baths.
		
		For instance,  at the EP of the photonic structure in Ref.~\cite{roccatiArxiv2021}, 
		the interaction between emitters (in the weak-coupling and Markovian regime) is described by an effective non-Hermitian Hamiltonian of the form
		\begin{equation}\label{atomHam}
		H_{\rm atoms}=-i\Gamma\sum_n \sigma_n^\dagger\sigma_n + i\Gamma \sum_n \sigma_{n+1}^\dagger\sigma_n \, ,
		\end{equation}
		with $\Gamma$ a rate that is quadratic in the emitter-cavity couping strength~\cite{roccatiArxiv2021}.
		This represents  an implementation of the (dissipative) fully non-reciprocal Hatano-Nelson model, cf.~Eq.~\eqref{HatanoPBCS}, 
		in which case $\sigma_n$ is the ladder operator of the $n$th emitter.

		Remarkably, it turns out that for an odd number of lattice cells/emitters, Hamiltonian~\eqref{atomHam}, which is translationally-invariant, is insensitive to the lattice boundary conditions. In particular, it arises even if the lattice is subject to open boundary conditions.

		\section{Conclusions}

		In this manuscript we discussed the relation between the effective description  through non-Hermitian Hamiltonians 
		and a  Markovian master equation for the full density matrix. 
		Non-Hermiticity is not a mere phenomenological tool to describe the system's openness, 
		but can exhibit under certain conditions non-trivial phenomena
		such as the appearance of exceptional points.
		
		We showed how non-Hermitian \PT-symmetric \textit{Hamiltonians} arise in the \textit{mean-field} description of open quadratic bosonic systems. The EP separates oscillating and amplifying dynamics for the mean-field, while beyond mean field it can be a critical point for quantum correlation (see Ref.~\cite{roccati2021quantum}).
		
		We outlined how non-Hermitian Hamiltonians, not necessarily \PT-~symmetric, can display counterintuitive  properties, and 
		how these can be connected to a Markovian master equation for the full density matrix.
		By allowing only certain cavities in a tight binding lattice to be lossy through a non-Hermitian description, it is possible to achieve non-reciprocal coupling between cavities. 
		

		\section*{Acknowledgments}
		
		We thank D.~Cilluffo for helpful comments. We acknowledge support from MIUR through project PRIN Project 2017SRN-BRK QUSHIP. FB acknowledges partial support from Palermo University and from G.N.F.M.~of the INdAM.

		\bibliographystyle{apsrev4-1}
		\bibliography{RPBCBIB}

\begin{thebibliography}{42}%
\makeatletter
\providecommand \@ifxundefined [1]{%
 \@ifx{#1\undefined}
}%
\providecommand \@ifnum [1]{%
 \ifnum #1\expandafter \@firstoftwo
 \else \expandafter \@secondoftwo
 \fi
}%
\providecommand \@ifx [1]{%
 \ifx #1\expandafter \@firstoftwo
 \else \expandafter \@secondoftwo
 \fi
}%
\providecommand \natexlab [1]{#1}%
\providecommand \enquote  [1]{``#1''}%
\providecommand \bibnamefont  [1]{#1}%
\providecommand \bibfnamefont [1]{#1}%
\providecommand \citenamefont [1]{#1}%
\providecommand \href@noop [0]{\@secondoftwo}%
\providecommand \href [0]{\begingroup \@sanitize@url \@href}%
\providecommand \@href[1]{\@@startlink{#1}\@@href}%
\providecommand \@@href[1]{\endgroup#1\@@endlink}%
\providecommand \@sanitize@url [0]{\catcode `\\12\catcode `\$12\catcode
  `\&12\catcode `\#12\catcode `\^12\catcode `\_12\catcode `\%12\relax}%
\providecommand \@@startlink[1]{}%
\providecommand \@@endlink[0]{}%
\providecommand \url  [0]{\begingroup\@sanitize@url \@url }%
\providecommand \@url [1]{\endgroup\@href {#1}{\urlprefix }}%
\providecommand \urlprefix  [0]{URL }%
\providecommand \Eprint [0]{\href }%
\providecommand \doibase [0]{http://dx.doi.org/}%
\providecommand \selectlanguage [0]{\@gobble}%
\providecommand \bibinfo  [0]{\@secondoftwo}%
\providecommand \bibfield  [0]{\@secondoftwo}%
\providecommand \translation [1]{[#1]}%
\providecommand \BibitemOpen [0]{}%
\providecommand \bibitemStop [0]{}%
\providecommand \bibitemNoStop [0]{.\EOS\space}%
\providecommand \EOS [0]{\spacefactor3000\relax}%
\providecommand \BibitemShut  [1]{\csname bibitem#1\endcsname}%
\let\auto@bib@innerbib\@empty
\bibitem [{\citenamefont {Cohen-Tannoudji}\ \emph {et~al.}(2019)\citenamefont
  {Cohen-Tannoudji}, \citenamefont {Diu},\ and\ \citenamefont
  {Lalo{\"e}}}]{cohen2019quantum}%
  \BibitemOpen
  \bibfield  {author} {\bibinfo {author} {\bibfnamefont {C.}~\bibnamefont
  {Cohen-Tannoudji}}, \bibinfo {author} {\bibfnamefont {B.}~\bibnamefont
  {Diu}}, \ and\ \bibinfo {author} {\bibfnamefont {F.}~\bibnamefont
  {Lalo{\"e}}},\ }\href@noop {} {\emph {\bibinfo {title} {Quantum Mechanics,
  Volume 1: Basic Concepts, Tools, and Applications}}}\ (\bibinfo  {publisher}
  {John Wiley \& Sons},\ \bibinfo {year} {2019})\BibitemShut {NoStop}%
\bibitem [{\citenamefont {Bender}\ \emph {et~al.}(2003)\citenamefont {Bender},
  \citenamefont {Brody},\ and\ \citenamefont {Jones}}]{benderAJoP2003}%
  \BibitemOpen
  \bibfield  {author} {\bibinfo {author} {\bibfnamefont {C.~M.}\ \bibnamefont
  {Bender}}, \bibinfo {author} {\bibfnamefont {D.~C.}\ \bibnamefont {Brody}}, \
  and\ \bibinfo {author} {\bibfnamefont {H.~F.}\ \bibnamefont {Jones}},\
  }\href@noop {} {\bibfield  {journal} {\bibinfo  {journal} {American Journal
  of Physics}\ }\textbf {\bibinfo {volume} {71}},\ \bibinfo {pages} {1095}
  (\bibinfo {year} {2003})}\BibitemShut {NoStop}%
\bibitem [{\citenamefont {Bender}(2007)}]{benderRPP2007}%
  \BibitemOpen
  \bibfield  {author} {\bibinfo {author} {\bibfnamefont {C.~M.}\ \bibnamefont
  {Bender}},\ }\href@noop {} {\bibfield  {journal} {\bibinfo  {journal}
  {Reports on Progress in Physics}\ }\textbf {\bibinfo {volume} {70}},\
  \bibinfo {pages} {947} (\bibinfo {year} {2007})}\BibitemShut {NoStop}%
\bibitem [{\citenamefont {Bergholtz}\ \emph {et~al.}(2021)\citenamefont
  {Bergholtz}, \citenamefont {Budich},\ and\ \citenamefont
  {Kunst}}]{BergholtzRMP2021}%
  \BibitemOpen
  \bibfield  {author} {\bibinfo {author} {\bibfnamefont {E.~J.}\ \bibnamefont
  {Bergholtz}}, \bibinfo {author} {\bibfnamefont {J.~C.}\ \bibnamefont
  {Budich}}, \ and\ \bibinfo {author} {\bibfnamefont {F.~K.}\ \bibnamefont
  {Kunst}},\ }\href@noop {} {\bibfield  {journal} {\bibinfo  {journal} {Reviews
  of Modern Physics}\ }\textbf {\bibinfo {volume} {93}},\ \bibinfo {pages}
  {015005} (\bibinfo {year} {2021})}\BibitemShut {NoStop}%
\bibitem [{\citenamefont {Kawabata}\ \emph {et~al.}(2019)\citenamefont
  {Kawabata}, \citenamefont {Shiozaki}, \citenamefont {Ueda},\ and\
  \citenamefont {Sato}}]{KawabataPRX2019}%
  \BibitemOpen
  \bibfield  {author} {\bibinfo {author} {\bibfnamefont {K.}~\bibnamefont
  {Kawabata}}, \bibinfo {author} {\bibfnamefont {K.}~\bibnamefont {Shiozaki}},
  \bibinfo {author} {\bibfnamefont {M.}~\bibnamefont {Ueda}}, \ and\ \bibinfo
  {author} {\bibfnamefont {M.}~\bibnamefont {Sato}},\ }\href@noop {} {\bibfield
   {journal} {\bibinfo  {journal} {Physical Review X}\ }\textbf {\bibinfo
  {volume} {9}},\ \bibinfo {pages} {041015} (\bibinfo {year}
  {2019})}\BibitemShut {NoStop}%
\bibitem [{\citenamefont {Ashida}\ \emph {et~al.}(2020)\citenamefont {Ashida},
  \citenamefont {Gong},\ and\ \citenamefont {Ueda}}]{AshidaAiP2020}%
  \BibitemOpen
  \bibfield  {author} {\bibinfo {author} {\bibfnamefont {Y.}~\bibnamefont
  {Ashida}}, \bibinfo {author} {\bibfnamefont {Z.}~\bibnamefont {Gong}}, \ and\
  \bibinfo {author} {\bibfnamefont {M.}~\bibnamefont {Ueda}},\ }\href@noop {}
  {\bibfield  {journal} {\bibinfo  {journal} {Advances in Physics}\ }\textbf
  {\bibinfo {volume} {69}},\ \bibinfo {pages} {249} (\bibinfo {year}
  {2020})}\BibitemShut {NoStop}%
\bibitem [{\citenamefont {Heiss}(2012)}]{heissJPAMT2012}%
  \BibitemOpen
  \bibfield  {author} {\bibinfo {author} {\bibfnamefont {W.}~\bibnamefont
  {Heiss}},\ }\href@noop {} {\bibfield  {journal} {\bibinfo  {journal} {Journal
  of Physics A: Mathematical and Theoretical}\ }\textbf {\bibinfo {volume}
  {45}},\ \bibinfo {pages} {444016} (\bibinfo {year} {2012})}\BibitemShut
  {NoStop}%
\bibitem [{\citenamefont {Miri}\ and\ \citenamefont
  {Alu}(2019)}]{Mirieaar7709}%
  \BibitemOpen
  \bibfield  {author} {\bibinfo {author} {\bibfnamefont {M.-A.}\ \bibnamefont
  {Miri}}\ and\ \bibinfo {author} {\bibfnamefont {A.}~\bibnamefont {Alu}},\
  }\href@noop {} {\bibfield  {journal} {\bibinfo  {journal} {Science}\ }\textbf
  {\bibinfo {volume} {363}} (\bibinfo {year} {2019})}\BibitemShut {NoStop}%
\bibitem [{\citenamefont {Bender}\ and\ \citenamefont
  {Boettcher}(1998)}]{benderPRL1998}%
  \BibitemOpen
  \bibfield  {author} {\bibinfo {author} {\bibfnamefont {C.~M.}\ \bibnamefont
  {Bender}}\ and\ \bibinfo {author} {\bibfnamefont {S.}~\bibnamefont
  {Boettcher}},\ }\href@noop {} {\bibfield  {journal} {\bibinfo  {journal}
  {Physical Review Letters}\ }\textbf {\bibinfo {volume} {80}},\ \bibinfo
  {pages} {5243} (\bibinfo {year} {1998})}\BibitemShut {NoStop}%
\bibitem [{\citenamefont {Brody}(2013)}]{brody2013biorthogonal}%
  \BibitemOpen
  \bibfield  {author} {\bibinfo {author} {\bibfnamefont {D.~C.}\ \bibnamefont
  {Brody}},\ }\href@noop {} {\bibfield  {journal} {\bibinfo  {journal} {Journal
  of Physics A: Mathematical and Theoretical}\ }\textbf {\bibinfo {volume}
  {47}},\ \bibinfo {pages} {035305} (\bibinfo {year} {2013})}\BibitemShut
  {NoStop}%
\bibitem [{\citenamefont {Mostafazadeh}(2010)}]{mostafazadeh2010conceptual}%
  \BibitemOpen
  \bibfield  {author} {\bibinfo {author} {\bibfnamefont {A.}~\bibnamefont
  {Mostafazadeh}},\ }\href@noop {} {\bibfield  {journal} {\bibinfo  {journal}
  {Physica Scripta}\ }\textbf {\bibinfo {volume} {82}},\ \bibinfo {pages}
  {038110} (\bibinfo {year} {2010})}\BibitemShut {NoStop}%
\bibitem [{\citenamefont {Bagarello}\ \emph {et~al.}(2015)\citenamefont
  {Bagarello}, \citenamefont {Gazeau}, \citenamefont {Szafraniec},\ and\
  \citenamefont {Znojil}}]{bagbook}%
  \BibitemOpen
  \bibfield  {author} {\bibinfo {author} {\bibfnamefont {F.}~\bibnamefont
  {Bagarello}}, \bibinfo {author} {\bibfnamefont {J.-P.}\ \bibnamefont
  {Gazeau}}, \bibinfo {author} {\bibfnamefont {F.~H.}\ \bibnamefont
  {Szafraniec}}, \ and\ \bibinfo {author} {\bibfnamefont {M.}~\bibnamefont
  {Znojil}},\ }\href@noop {} {\emph {\bibinfo {title} {Non-selfadjoint
  operators in quantum physics: Mathematical aspects}}}\ (\bibinfo  {publisher}
  {John Wiley \& Sons},\ \bibinfo {year} {2015})\BibitemShut {NoStop}%
\bibitem [{\citenamefont {Giordanelli}\ and\ \citenamefont
  {Graf}(2013)}]{giordanelli2013real}%
  \BibitemOpen
  \bibfield  {author} {\bibinfo {author} {\bibfnamefont {I.}~\bibnamefont
  {Giordanelli}}\ and\ \bibinfo {author} {\bibfnamefont {G.~M.}\ \bibnamefont
  {Graf}},\ }\href@noop {} {\bibfield  {journal} {\bibinfo  {journal} {arXiv
  preprint arXiv:1310.7767}\ } (\bibinfo {year} {2013})}\BibitemShut {NoStop}%
\bibitem [{\citenamefont {El-Ganainy}\ \emph {et~al.}(2018)\citenamefont
  {El-Ganainy}, \citenamefont {Makris}, \citenamefont {Khajavikhan},
  \citenamefont {Musslimani}, \citenamefont {Rotter},\ and\ \citenamefont
  {Christodoulides}}]{el-ganainyNP2018}%
  \BibitemOpen
  \bibfield  {author} {\bibinfo {author} {\bibfnamefont {R.}~\bibnamefont
  {El-Ganainy}}, \bibinfo {author} {\bibfnamefont {K.~G.}\ \bibnamefont
  {Makris}}, \bibinfo {author} {\bibfnamefont {M.}~\bibnamefont {Khajavikhan}},
  \bibinfo {author} {\bibfnamefont {Z.~H.}\ \bibnamefont {Musslimani}},
  \bibinfo {author} {\bibfnamefont {S.}~\bibnamefont {Rotter}}, \ and\ \bibinfo
  {author} {\bibfnamefont {D.~N.}\ \bibnamefont {Christodoulides}},\
  }\href@noop {} {\bibfield  {journal} {\bibinfo  {journal} {Nature Physics}\
  }\textbf {\bibinfo {volume} {14}},\ \bibinfo {pages} {11} (\bibinfo {year}
  {2018})}\BibitemShut {NoStop}%
\bibitem [{\citenamefont {R{\"u}ter}\ \emph {et~al.}(2010)\citenamefont
  {R{\"u}ter}, \citenamefont {Makris}, \citenamefont {El-Ganainy},
  \citenamefont {Christodoulides}, \citenamefont {Segev},\ and\ \citenamefont
  {Kip}}]{ruterNP2010}%
  \BibitemOpen
  \bibfield  {author} {\bibinfo {author} {\bibfnamefont {C.~E.}\ \bibnamefont
  {R{\"u}ter}}, \bibinfo {author} {\bibfnamefont {K.~G.}\ \bibnamefont
  {Makris}}, \bibinfo {author} {\bibfnamefont {R.}~\bibnamefont {El-Ganainy}},
  \bibinfo {author} {\bibfnamefont {D.~N.}\ \bibnamefont {Christodoulides}},
  \bibinfo {author} {\bibfnamefont {M.}~\bibnamefont {Segev}}, \ and\ \bibinfo
  {author} {\bibfnamefont {D.}~\bibnamefont {Kip}},\ }\href@noop {} {\bibfield
  {journal} {\bibinfo  {journal} {Nature physics}\ }\textbf {\bibinfo {volume}
  {6}},\ \bibinfo {pages} {192} (\bibinfo {year} {2010})}\BibitemShut {NoStop}%
\bibitem [{\citenamefont {Bender}(2019)}]{bender2019pt}%
  \BibitemOpen
  \bibfield  {author} {\bibinfo {author} {\bibfnamefont {C.~M.}\ \bibnamefont
  {Bender}},\ }\href@noop {} {\emph {\bibinfo {title} {PT symmetry: In quantum
  and classical physics}}}\ (\bibinfo  {publisher} {World Scientific},\
  \bibinfo {year} {2019})\BibitemShut {NoStop}%
\bibitem [{\citenamefont {Wiersig}(2020)}]{WiersigPR2020}%
  \BibitemOpen
  \bibfield  {author} {\bibinfo {author} {\bibfnamefont {J.}~\bibnamefont
  {Wiersig}},\ }\href@noop {} {\bibfield  {journal} {\bibinfo  {journal}
  {Photonics Research}\ }\textbf {\bibinfo {volume} {8}},\ \bibinfo {pages}
  {1457} (\bibinfo {year} {2020})}\BibitemShut {NoStop}%
\bibitem [{\citenamefont {Roccati}\ \emph
  {et~al.}(2021{\natexlab{a}})\citenamefont {Roccati}, \citenamefont {Lorenzo},
  \citenamefont {Palma}, \citenamefont {Landi}, \citenamefont {Brunelli},\ and\
  \citenamefont {Ciccarello}}]{roccati2021quantum}%
  \BibitemOpen
  \bibfield  {author} {\bibinfo {author} {\bibfnamefont {F.}~\bibnamefont
  {Roccati}}, \bibinfo {author} {\bibfnamefont {S.}~\bibnamefont {Lorenzo}},
  \bibinfo {author} {\bibfnamefont {G.~M.}\ \bibnamefont {Palma}}, \bibinfo
  {author} {\bibfnamefont {G.~T.}\ \bibnamefont {Landi}}, \bibinfo {author}
  {\bibfnamefont {M.}~\bibnamefont {Brunelli}}, \ and\ \bibinfo {author}
  {\bibfnamefont {F.}~\bibnamefont {Ciccarello}},\ }\href@noop {} {\bibfield
  {journal} {\bibinfo  {journal} {Quantum Science and Technology}\ }\textbf
  {\bibinfo {volume} {6}},\ \bibinfo {pages} {025005} (\bibinfo {year}
  {2021}{\natexlab{a}})}\BibitemShut {NoStop}%
\bibitem [{\citenamefont {Purkayastha}\ \emph {et~al.}(2020)\citenamefont
  {Purkayastha}, \citenamefont {Kulkarni},\ and\ \citenamefont
  {Joglekar}}]{PurkayasthaPRR2020}%
  \BibitemOpen
  \bibfield  {author} {\bibinfo {author} {\bibfnamefont {A.}~\bibnamefont
  {Purkayastha}}, \bibinfo {author} {\bibfnamefont {M.}~\bibnamefont
  {Kulkarni}}, \ and\ \bibinfo {author} {\bibfnamefont {Y.~N.}\ \bibnamefont
  {Joglekar}},\ }\href@noop {} {\bibfield  {journal} {\bibinfo  {journal}
  {Physical Review Research}\ }\textbf {\bibinfo {volume} {2}},\ \bibinfo
  {pages} {043075} (\bibinfo {year} {2020})}\BibitemShut {NoStop}%
\bibitem [{\citenamefont {Ornigotti}\ and\ \citenamefont
  {Szameit}(2014)}]{Ornigotti_2014}%
  \BibitemOpen
  \bibfield  {author} {\bibinfo {author} {\bibfnamefont {M.}~\bibnamefont
  {Ornigotti}}\ and\ \bibinfo {author} {\bibfnamefont {A.}~\bibnamefont
  {Szameit}},\ }\href@noop {} {\bibfield  {journal} {\bibinfo  {journal}
  {Journal of Optics}\ }\textbf {\bibinfo {volume} {16}},\ \bibinfo {pages}
  {065501} (\bibinfo {year} {2014})}\BibitemShut {NoStop}%
\bibitem [{\citenamefont {Joglekar}\ and\ \citenamefont
  {Harter}(2018)}]{JoglekarPRes2018}%
  \BibitemOpen
  \bibfield  {author} {\bibinfo {author} {\bibfnamefont {Y.~N.}\ \bibnamefont
  {Joglekar}}\ and\ \bibinfo {author} {\bibfnamefont {A.~K.}\ \bibnamefont
  {Harter}},\ }\href@noop {} {\bibfield  {journal} {\bibinfo  {journal}
  {Photonics Research}\ }\textbf {\bibinfo {volume} {6}},\ \bibinfo {pages}
  {A51} (\bibinfo {year} {2018})}\BibitemShut {NoStop}%
\bibitem [{\citenamefont {Breuer}\ \emph {et~al.}(2002)\citenamefont {Breuer},
  \citenamefont {Petruccione} \emph {et~al.}}]{breuer2007}%
  \BibitemOpen
  \bibfield  {author} {\bibinfo {author} {\bibfnamefont {H.-P.}\ \bibnamefont
  {Breuer}}, \bibinfo {author} {\bibfnamefont {F.}~\bibnamefont {Petruccione}},
   \emph {et~al.},\ }\href@noop {} {\emph {\bibinfo {title} {The theory of open
  quantum systems}}}\ (\bibinfo  {publisher} {Oxford University Press on
  Demand},\ \bibinfo {year} {2002})\BibitemShut {NoStop}%
\bibitem [{\citenamefont {Rivas}\ and\ \citenamefont
  {Huelga}(2012)}]{rivas2012open}%
  \BibitemOpen
  \bibfield  {author} {\bibinfo {author} {\bibfnamefont {A.}~\bibnamefont
  {Rivas}}\ and\ \bibinfo {author} {\bibfnamefont {S.~F.}\ \bibnamefont
  {Huelga}},\ }\href@noop {} {\emph {\bibinfo {title} {Open quantum
  systems}}},\ Vol.~\bibinfo {volume} {10}\ (\bibinfo  {publisher} {Springer},\
  \bibinfo {year} {2012})\BibitemShut {NoStop}%
\bibitem [{\citenamefont {Ciccarello}\ \emph {et~al.}(2021)\citenamefont
  {Ciccarello}, \citenamefont {Lorenzo}, \citenamefont {Giovannetti},\ and\
  \citenamefont {Palma}}]{ciccarello2021quantum}%
  \BibitemOpen
  \bibfield  {author} {\bibinfo {author} {\bibfnamefont {F.}~\bibnamefont
  {Ciccarello}}, \bibinfo {author} {\bibfnamefont {S.}~\bibnamefont {Lorenzo}},
  \bibinfo {author} {\bibfnamefont {V.}~\bibnamefont {Giovannetti}}, \ and\
  \bibinfo {author} {\bibfnamefont {G.~M.}\ \bibnamefont {Palma}},\ }\href@noop
  {} {\bibfield  {journal} {\bibinfo  {journal} {arXiv preprint
  arXiv:2106.11974}\ } (\bibinfo {year} {2021})}\BibitemShut {NoStop}%
\bibitem [{\citenamefont {Hall}(2013)}]{hall2013}%
  \BibitemOpen
  \bibfield  {author} {\bibinfo {author} {\bibfnamefont {B.~C.}\ \bibnamefont
  {Hall}},\ }\href@noop {} {\emph {\bibinfo {title} {Quantum theory for
  mathematicians}}},\ Vol.\ \bibinfo {volume} {267}\ (\bibinfo  {publisher}
  {Springer},\ \bibinfo {year} {2013})\BibitemShut {NoStop}%
\bibitem [{\citenamefont {Landau}(1927)}]{landau1965}%
  \BibitemOpen
  \bibfield  {author} {\bibinfo {author} {\bibfnamefont {L.}~\bibnamefont
  {Landau}},\ }\href@noop {} {\bibfield  {journal} {\bibinfo  {journal} {Z.
  Phys}\ }\textbf {\bibinfo {volume} {45}},\ \bibinfo {pages} {430} (\bibinfo
  {year} {1927})}\BibitemShut {NoStop}%
\bibitem [{\citenamefont {Cohen-Tannoudji}\ \emph {et~al.}(1993)\citenamefont
  {Cohen-Tannoudji}, \citenamefont {Dupont-Roc},\ and\ \citenamefont
  {Grynberg}}]{CTap1992}%
  \BibitemOpen
  \bibfield  {author} {\bibinfo {author} {\bibfnamefont {C.}~\bibnamefont
  {Cohen-Tannoudji}}, \bibinfo {author} {\bibfnamefont {J.}~\bibnamefont
  {Dupont-Roc}}, \ and\ \bibinfo {author} {\bibfnamefont {G.}~\bibnamefont
  {Grynberg}},\ }\href@noop {} {\enquote {\bibinfo {title} {Atom--photon
  interactions: Basic processes and applications},}\ } (\bibinfo {year}
  {1993})\BibitemShut {NoStop}%
\bibitem [{\citenamefont {Minganti}\ \emph {et~al.}(2019)\citenamefont
  {Minganti}, \citenamefont {Miranowicz}, \citenamefont {Chhajlany},\ and\
  \citenamefont {Nori}}]{mingantiPRA2019}%
  \BibitemOpen
  \bibfield  {author} {\bibinfo {author} {\bibfnamefont {F.}~\bibnamefont
  {Minganti}}, \bibinfo {author} {\bibfnamefont {A.}~\bibnamefont
  {Miranowicz}}, \bibinfo {author} {\bibfnamefont {R.~W.}\ \bibnamefont
  {Chhajlany}}, \ and\ \bibinfo {author} {\bibfnamefont {F.}~\bibnamefont
  {Nori}},\ }\href@noop {} {\bibfield  {journal} {\bibinfo  {journal} {Physical
  Review A}\ }\textbf {\bibinfo {volume} {100}},\ \bibinfo {pages} {062131}
  (\bibinfo {year} {2019})}\BibitemShut {NoStop}%
\bibitem [{\citenamefont {Olivares}(2012)}]{olivares2012quantum}%
  \BibitemOpen
  \bibfield  {author} {\bibinfo {author} {\bibfnamefont {S.}~\bibnamefont
  {Olivares}},\ }\href@noop {} {\bibfield  {journal} {\bibinfo  {journal} {The
  European Physical Journal Special Topics}\ }\textbf {\bibinfo {volume}
  {203}},\ \bibinfo {pages} {3} (\bibinfo {year} {2012})}\BibitemShut {NoStop}%
\bibitem [{\citenamefont {Gardiner}\ \emph {et~al.}(2004)\citenamefont
  {Gardiner}, \citenamefont {Zoller},\ and\ \citenamefont
  {Zoller}}]{gardiner2004}%
  \BibitemOpen
  \bibfield  {author} {\bibinfo {author} {\bibfnamefont {C.}~\bibnamefont
  {Gardiner}}, \bibinfo {author} {\bibfnamefont {P.}~\bibnamefont {Zoller}}, \
  and\ \bibinfo {author} {\bibfnamefont {P.}~\bibnamefont {Zoller}},\
  }\href@noop {} {\emph {\bibinfo {title} {Quantum noise: a handbook of
  Markovian and non-Markovian quantum stochastic methods with applications to
  quantum optics}}}\ (\bibinfo  {publisher} {Springer Science \& Business
  Media},\ \bibinfo {year} {2004})\BibitemShut {NoStop}%
\bibitem [{\citenamefont {Peng}\ \emph {et~al.}(2014)\citenamefont {Peng},
  \citenamefont {{\"O}zdemir}, \citenamefont {Lei}, \citenamefont {Monifi},
  \citenamefont {Gianfreda}, \citenamefont {Long}, \citenamefont {Fan},
  \citenamefont {Nori}, \citenamefont {Bender},\ and\ \citenamefont
  {Yang}}]{pengNP2014}%
  \BibitemOpen
  \bibfield  {author} {\bibinfo {author} {\bibfnamefont {B.}~\bibnamefont
  {Peng}}, \bibinfo {author} {\bibfnamefont {{\c{S}}.~K.}\ \bibnamefont
  {{\"O}zdemir}}, \bibinfo {author} {\bibfnamefont {F.}~\bibnamefont {Lei}},
  \bibinfo {author} {\bibfnamefont {F.}~\bibnamefont {Monifi}}, \bibinfo
  {author} {\bibfnamefont {M.}~\bibnamefont {Gianfreda}}, \bibinfo {author}
  {\bibfnamefont {G.~L.}\ \bibnamefont {Long}}, \bibinfo {author}
  {\bibfnamefont {S.}~\bibnamefont {Fan}}, \bibinfo {author} {\bibfnamefont
  {F.}~\bibnamefont {Nori}}, \bibinfo {author} {\bibfnamefont {C.~M.}\
  \bibnamefont {Bender}}, \ and\ \bibinfo {author} {\bibfnamefont
  {L.}~\bibnamefont {Yang}},\ }\href@noop {} {\bibfield  {journal} {\bibinfo
  {journal} {Nature Physics}\ }\textbf {\bibinfo {volume} {10}},\ \bibinfo
  {pages} {394} (\bibinfo {year} {2014})}\BibitemShut {NoStop}%
\bibitem [{\citenamefont {Dast}\ \emph {et~al.}(2014)\citenamefont {Dast},
  \citenamefont {Haag}, \citenamefont {Cartarius},\ and\ \citenamefont
  {Wunner}}]{dastPRA2014}%
  \BibitemOpen
  \bibfield  {author} {\bibinfo {author} {\bibfnamefont {D.}~\bibnamefont
  {Dast}}, \bibinfo {author} {\bibfnamefont {D.}~\bibnamefont {Haag}}, \bibinfo
  {author} {\bibfnamefont {H.}~\bibnamefont {Cartarius}}, \ and\ \bibinfo
  {author} {\bibfnamefont {G.}~\bibnamefont {Wunner}},\ }\href@noop {}
  {\bibfield  {journal} {\bibinfo  {journal} {Physical Review A}\ }\textbf
  {\bibinfo {volume} {90}},\ \bibinfo {pages} {052120} (\bibinfo {year}
  {2014})}\BibitemShut {NoStop}%
\bibitem [{\citenamefont {Longhi}(2019)}]{longhi2019quantum}%
  \BibitemOpen
  \bibfield  {author} {\bibinfo {author} {\bibfnamefont {S.}~\bibnamefont
  {Longhi}},\ }\href@noop {} {\bibfield  {journal} {\bibinfo  {journal}
  {Physical Review A}\ }\textbf {\bibinfo {volume} {100}},\ \bibinfo {pages}
  {022123} (\bibinfo {year} {2019})}\BibitemShut {NoStop}%
\bibitem [{\citenamefont {Lieu}(2018)}]{LieuPRB2018}%
  \BibitemOpen
  \bibfield  {author} {\bibinfo {author} {\bibfnamefont {S.}~\bibnamefont
  {Lieu}},\ }\href@noop {} {\bibfield  {journal} {\bibinfo  {journal} {Physical
  Review B}\ }\textbf {\bibinfo {volume} {97}},\ \bibinfo {pages} {045106}
  (\bibinfo {year} {2018})}\BibitemShut {NoStop}%
\bibitem [{\citenamefont {Hatano}\ and\ \citenamefont
  {Nelson}(1996)}]{HatanoPRL96}%
  \BibitemOpen
  \bibfield  {author} {\bibinfo {author} {\bibfnamefont {N.}~\bibnamefont
  {Hatano}}\ and\ \bibinfo {author} {\bibfnamefont {D.~R.}\ \bibnamefont
  {Nelson}},\ }\href@noop {} {\bibfield  {journal} {\bibinfo  {journal}
  {Physical review letters}\ }\textbf {\bibinfo {volume} {77}},\ \bibinfo
  {pages} {570} (\bibinfo {year} {1996})}\BibitemShut {NoStop}%
\bibitem [{\citenamefont {Wang}\ \emph {et~al.}(2021)\citenamefont {Wang},
  \citenamefont {Dutt}, \citenamefont {Yang}, \citenamefont {Wojcik},
  \citenamefont {Vu{\v{c}}kovi{\'c}},\ and\ \citenamefont
  {Fan}}]{WangSCIENCE2021}%
  \BibitemOpen
  \bibfield  {author} {\bibinfo {author} {\bibfnamefont {K.}~\bibnamefont
  {Wang}}, \bibinfo {author} {\bibfnamefont {A.}~\bibnamefont {Dutt}}, \bibinfo
  {author} {\bibfnamefont {K.~Y.}\ \bibnamefont {Yang}}, \bibinfo {author}
  {\bibfnamefont {C.~C.}\ \bibnamefont {Wojcik}}, \bibinfo {author}
  {\bibfnamefont {J.}~\bibnamefont {Vu{\v{c}}kovi{\'c}}}, \ and\ \bibinfo
  {author} {\bibfnamefont {S.}~\bibnamefont {Fan}},\ }\href@noop {} {\bibfield
  {journal} {\bibinfo  {journal} {Science}\ }\textbf {\bibinfo {volume}
  {371}},\ \bibinfo {pages} {1240} (\bibinfo {year} {2021})}\BibitemShut
  {NoStop}%
\bibitem [{\citenamefont {Roccati}(2021)}]{RoccatiPRA2021}%
  \BibitemOpen
  \bibfield  {author} {\bibinfo {author} {\bibfnamefont {F.}~\bibnamefont
  {Roccati}},\ }\href@noop {} {\bibfield  {journal} {\bibinfo  {journal} {Phys.
  Rev. A}\ }\textbf {\bibinfo {volume} {104}},\ \bibinfo {pages} {022215}
  (\bibinfo {year} {2021})}\BibitemShut {NoStop}%
\bibitem [{\citenamefont {Alvarez}\ \emph {et~al.}(2018)\citenamefont
  {Alvarez}, \citenamefont {Vargas}, \citenamefont {Berdakin},\ and\
  \citenamefont {Torres}}]{alvarez2018topological}%
  \BibitemOpen
  \bibfield  {author} {\bibinfo {author} {\bibfnamefont {V.~M.}\ \bibnamefont
  {Alvarez}}, \bibinfo {author} {\bibfnamefont {J.~B.}\ \bibnamefont {Vargas}},
  \bibinfo {author} {\bibfnamefont {M.}~\bibnamefont {Berdakin}}, \ and\
  \bibinfo {author} {\bibfnamefont {L.~F.}\ \bibnamefont {Torres}},\
  }\href@noop {} {\bibfield  {journal} {\bibinfo  {journal} {The European
  Physical Journal Special Topics}\ }\textbf {\bibinfo {volume} {227}},\
  \bibinfo {pages} {1295} (\bibinfo {year} {2018})}\BibitemShut {NoStop}%
\bibitem [{\citenamefont {Longhi}\ \emph {et~al.}(2015)\citenamefont {Longhi},
  \citenamefont {Gatti},\ and\ \citenamefont {Della~Valle}}]{LonghiSciRep2015}%
  \BibitemOpen
  \bibfield  {author} {\bibinfo {author} {\bibfnamefont {S.}~\bibnamefont
  {Longhi}}, \bibinfo {author} {\bibfnamefont {D.}~\bibnamefont {Gatti}}, \
  and\ \bibinfo {author} {\bibfnamefont {G.}~\bibnamefont {Della~Valle}},\
  }\href@noop {} {\bibfield  {journal} {\bibinfo  {journal} {Scientific
  reports}\ }\textbf {\bibinfo {volume} {5}},\ \bibinfo {pages} {1} (\bibinfo
  {year} {2015})}\BibitemShut {NoStop}%
\bibitem [{\citenamefont {Metelmann}\ and\ \citenamefont
  {Clerk}(2015)}]{MetelmannPRX2015}%
  \BibitemOpen
  \bibfield  {author} {\bibinfo {author} {\bibfnamefont {A.}~\bibnamefont
  {Metelmann}}\ and\ \bibinfo {author} {\bibfnamefont {A.~A.}\ \bibnamefont
  {Clerk}},\ }\href@noop {} {\bibfield  {journal} {\bibinfo  {journal}
  {Physical Review X}\ }\textbf {\bibinfo {volume} {5}},\ \bibinfo {pages}
  {021025} (\bibinfo {year} {2015})}\BibitemShut {NoStop}%
\bibitem [{\citenamefont {Clerk}(2022)}]{ClerkArxiv2022}%
  \BibitemOpen
  \bibfield  {author} {\bibinfo {author} {\bibfnamefont {A.~A.}\ \bibnamefont
  {Clerk}},\ }\href@noop {} {\bibfield  {journal} {\bibinfo  {journal} {arXiv
  preprint arXiv:2201.00894}\ } (\bibinfo {year} {2022})}\BibitemShut {NoStop}%
\bibitem [{\citenamefont {Roccati}\ \emph
  {et~al.}(2021{\natexlab{b}})\citenamefont {Roccati}, \citenamefont {Lorenzo},
  \citenamefont {Calaj{\`o}}, \citenamefont {Palma}, \citenamefont {Carollo},\
  and\ \citenamefont {Ciccarello}}]{roccatiArxiv2021}%
  \BibitemOpen
  \bibfield  {author} {\bibinfo {author} {\bibfnamefont {F.}~\bibnamefont
  {Roccati}}, \bibinfo {author} {\bibfnamefont {S.}~\bibnamefont {Lorenzo}},
  \bibinfo {author} {\bibfnamefont {G.}~\bibnamefont {Calaj{\`o}}}, \bibinfo
  {author} {\bibfnamefont {G.~M.}\ \bibnamefont {Palma}}, \bibinfo {author}
  {\bibfnamefont {A.}~\bibnamefont {Carollo}}, \ and\ \bibinfo {author}
  {\bibfnamefont {F.}~\bibnamefont {Ciccarello}},\ }\href@noop {} {\bibfield
  {journal} {\bibinfo  {journal} {arXiv preprint arXiv:2109.13255}\ } (\bibinfo
  {year} {2021}{\natexlab{b}})}\BibitemShut {NoStop}%
\end{thebibliography}%

\end{document}